\documentclass{article}

\usepackage{arxiv}

\usepackage[utf8]{inputenc} 
\usepackage[T1]{fontenc}    
\usepackage{hyperref}       
\usepackage{url}            
\usepackage{booktabs}       
\usepackage{amsfonts}       
\usepackage{nicefrac}       
\usepackage{microtype}      
\usepackage{lipsum}
\usepackage{framed,multirow}

\usepackage{amssymb}
\usepackage{latexsym}

\usepackage{url}
\usepackage{flushend}

\providecommand{\keywords}[1]{\textbf{\textit{Index terms---}} #1}

\usepackage{amsmath}
\usepackage{amssymb}
\usepackage{paralist}
\usepackage{enumitem}
\usepackage{mathtools}

\usepackage{multirow}
\usepackage{tabularx}
\usepackage{cuted}
\usepackage{capt-of}
\usepackage{ctable}

\title{Degenerative Adversarial NeuroImage Nets for Brain Scan Simulations: Application in Ageing and Dementia}

\author{Daniele Ravi\\
  Centre for Medical Image Computing (CMIC)\\
  Department of Computer Science\\
  University College London \\ 
  London - United Kingdom \\
  \texttt{d.ravi@ucl.ac.uk} \\
  \and
  \textbf{Stefano B.Blumberg}\\
  Centre for Medical Image Computing (CMIC)\\
  Department of Computer Science\\
  University College London\\ 
  London - United Kingdom
  \and
  \textbf{Silvia Ingala}\\
  Department of Radiology and Nuclear Medicine\\
  Neuroscience Campus Amsterdam\\
  VU University Medical Center\\ 
  Amsterdam - Netherlands\\
  \and
  \textbf{Frederik Barkhof}\\
  Insititutes of Neurology and Healthcare Engineering\\
  University College London\\
  London - United Kingdom\\
  \and
  \textbf{Daniel C. Alexander}\\
  Centre for Medical Image Computing (CMIC)\\
  Department of Computer Science\\
  University College London \\
  London - United Kingdom\\
  \and
  \textbf{Neil P. Oxtoby}\\
  Centre for Medical Image Computing (CMIC)\\
  Department of Computer Science\\
  University College London \\
  London - United Kingdom
  \and
  \\
\textbf{for the Alzheimer's Disease Neuroimaging Initiative}\thanks{Data used in preparation of this article were obtained from the Alzheimer's Disease Neuroimaging Initiative (ADNI) database (adni.loni.usc.edu). As such, the investigators within the ADNI contributed to the design and implementation of ADNI and/or provided data but did not participate in analysis or writing of this report.
A complete listing of ADNI investigators can be found at:
\url{http://adni.loni.usc.edu/wp-content/uploads/how_to_apply/ADNI_Acknowledgement_List.pdf}} \\

}

\begin{document}
\maketitle

\begin{abstract}
Accurate and realistic simulation of high-dimensional medical images has become an important research area relevant to many AI-enabled healthcare applications. However, current state-of-the-art approaches lack the ability to produce satisfactory high-resolution and accurate subject-specific images. In this work, we present a deep learning framework, namely 4D-Degenerative Adversarial NeuroImage Net (4D-DANI-Net), to generate high-resolution, longitudinal MRI scans that mimic subject-specific neurodegeneration in ageing and dementia. 4D-DANI-Net is a modular framework based on adversarial training and a set of novel spatiotemporal, biologically-informed constraints. To ensure efficient training and overcome memory limitations affecting such high-dimensional problems, we rely on three key technological advances: i) a new 3D training consistency mechanism called Profile Weight Functions (PWFs), ii) a 3D super-resolution module and iii) a transfer learning strategy to fine-tune the system for a given individual. To evaluate our approach, we trained the framework on 9852 T1-weighted MRI scans from 876 participants in the Alzheimer's Disease Neuroimaging Initiative dataset and held out a separate test set of 1283 MRI scans from 170 participants for quantitative and qualitative assessment of the personalised time series of synthetic images. We performed three evaluations: i) image quality assessment; ii) quantifying the accuracy of regional brain volumes over and above benchmark models; and iii) quantifying visual perception of the synthetic images by medical experts. Overall, both quantitative and qualitative results show that 4D-DANI-Net produces realistic, low-artefact, personalised time series of synthetic T1 MRI that outperforms benchmark models. 
\end{abstract}



\section{Introduction}
The increasing availability of big data in healthcare and medicine has produced a boom in AI-enabled healthcare tools, particularly in medical image analysis. However, in various contexts of this research area, there is a lack of ground truth data, which presents challenges for trust and reliability of the related AI-based tools. Therefore, medical image simulation able to generate accurate and realistic data for model validation, can be a vital ingredient in the development of these new technologies. Such simulators are also important for data augmentation when training AI data-hungry models in situations where insufficient samples are available, e.g., in rarer diseases, or to recover missing images in longitudinal studies and predict future disease courses (virtual placebo). Here we introduce a novel computationally efficient 4D brain image simulation approach and demonstrate its capabilities in a neuroimaging application. 

Neurodegenerative diseases are a major challenge of 21st-century medicine, with the increasing incidence of these age-related diseases expected to continue to rise as the global population ages. This has inspired an explosion in medical data-sharing initiatives including from healthcare records (e.g., Alzheimer’s Disease Data Initiative (ADDI)), and large observational research studies such as the Alzheimer's Disease Neuroimaging Initiative (ADNI). Neuroimaging, such as Magnetic Resonance Imaging (MRI), is able to probe neurodegenerative diseases noninvasively and has provided well-established biomarkers for tracking disease progression in the clinic~\cite{frisoni2010clinical}. 
The data-sharing revolution has inspired the development of a suite of data-driven computational modelling methods for understanding and predicting disease progression~\cite{imaging_plus_x,golriz2020challenges}, with imaging playing a key role. Despite these efforts, suitable medical image simulators are relatively few and lagging behind, because simulating realistic and accurate neuroimaging data presents multiple challenges both biologically and computationally, many of which we address in this work.

Here, we introduce 4D-DANI-Net: a computationally-efficient framework for synthesizing a realistic, accurate, and personalized time series of high-resolution brain images for an individual conditioned on disease stage (clinical diagnosis) and age. 

Our contributions can be summarized as follows: i) we designed a new pipeline that enables the simulation of 4D MRI in both ageing and disease; ii) we proposed a sequence of memory-efficient techniques designed to improve training stability, reduce image artefacts, and increase individualization; and iii) we proposed a new validation protocol based on volumetric comparison to assess the accuracy of such a system.

We demonstrate our framework in the context of Alzheimer's disease and our experiments extensively analyze the capabilities of 4D-DANI-Net through quantitative and qualitative assessment, after training on a large dataset consisting of 9652 T1-weighted MRI from the ADNI and validate on a separate test set of 1216 MRI (also from the ADNI).

The paper is structured as follows: in Section~\ref{background}, we describe relevant previous work; in Section~\ref{sec:baseline}, we summarize our new framework; in Section~\ref{sec:training}, we describe the data set and our training protocol. Experimental results are presented in Section~\ref{sec:result}, and we conclude in Section~\ref{sec:conclusion}.

\section{Background}\label{background}

Computational disease progression modelling is a discipline that studies biophysical mechanisms and observable patterns of pathology spread and symptoms in chronic diseases. Such models are motivated by one or more applications including predicting the future course and providing insight for disease staging, which could help to achieve early diagnosis and personalized care. For a review of data-driven disease progression models, see~\cite{imaging_plus_x}. Briefly, the input to many disease progression models~\cite{fonteijn2012event,young2014data,jedynak2012computational,donohue2014estimating,lorenzi2019probabilistic,oxtoby2018data,young2018uncovering} is unstructured data such as scalar biomarkers, including those extracted from MRI for assessing neurodegeneration. Spatiotemporal models, e.g.,~\cite{lorenzi2015disentangling,durrleman2013toward}, attempt to incorporate structural information from the MRI themselves. All these models aim to produce quantitative templates of disease progression that promise utility for, e.g., recruiting the right patients at the right time into clinical trials. An MRI simulator has a key role to play in validating such models for these important applications. \textcolor{black}{Other potential clinical applications include enhancing AI interpretability by providing counterfactual visual examples that help humans identify errors in classifications made by AI systems to understand how marginal decisions come about~\cite{goyal2019counterfactual,woods2019adversarial,chang2021explaining}. Lastly, such simulators can be used to augment medical imaging datasets by creating new realistic samples required to train data-intensive AI algorithms when data collection is infeasible or too expensive~\cite{ravi2019adversarial,Prakosa6310066,chen2021synthetic}}.

Current MRI simulators can be divided into two categories:

i) biomechanical/physics-based models which describe the brain deformations in mechanical terms such as strain, displacement and stress. These models consider geometry, boundary conditions, loading, and material properties in their definition~\cite{miller2019biomechanical,khanal2017simulating};

ii) data-driven/learning-based models capable of understanding and predicting disease progression. These approaches often use machine learning, including deep learning techniques to distil information from big data~\cite{ravi2016deep}. Among these, a type of neural network that is particularly useful for generative modelling and simulation is the Generative Adversarial Network (GAN)~\cite{goodfellow2014generative}, which can generate new samples that plausibly come from an existing distribution of real data. To do this, GANs are trained using two neural network models: a generator that learns to generate new plausible samples, and a discriminator that learns to differentiate generated examples from real examples. However, due to the high spatial dimensionality (many voxels per scan) and temporal sparsity of MRI data (few time-points per individual), training such type of networks is challenging and computationally expensive.

In particular, current MRI simulators suffer three key limitations that severely limit their utility: i) lack of individualization; ii) poor image resolution; iii) limited to 2D images.

\begin{figure*}[t]
\begin{center}
\includegraphics[width=1\textwidth,trim={0cm 0cm 3.5cm 0cm},clip]{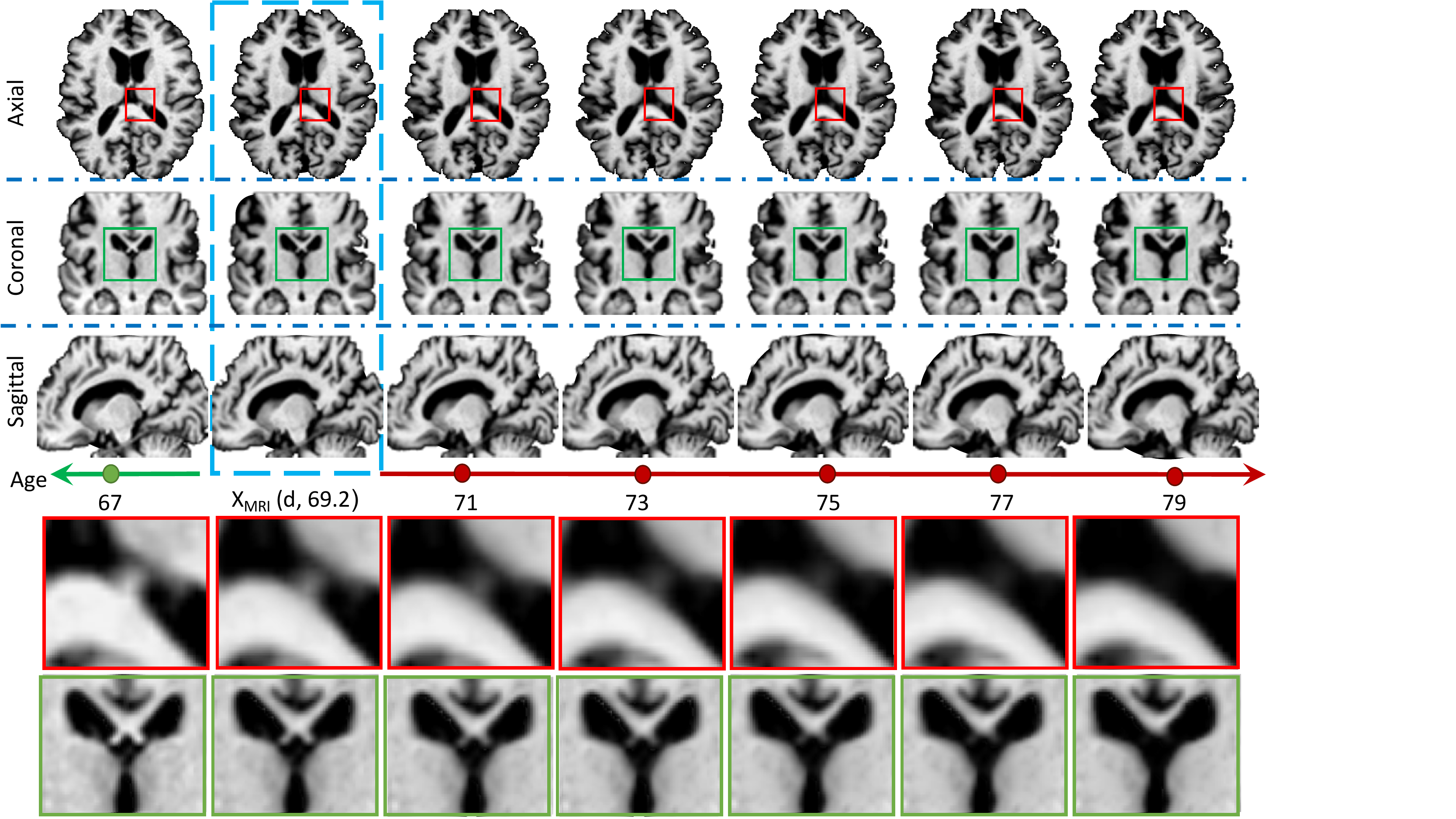}
\caption{\textcolor{black}{This figure shows in a 3-plane orientation, the longitudinal MRI synthesized using our approach for a CN subject at the age of 69.2. The blue box is the input MRI, all the other are our synthesized MRI scans. Two magnified regions are reported at the bottom of the figure.}
\label{fig:progression}}
\end{center}
\end{figure*}

\begin{figure*}[t]
\begin{center}
    \includegraphics[width=0.9\textwidth,trim={0cm 1.55cm 13cm 0cm},clip]{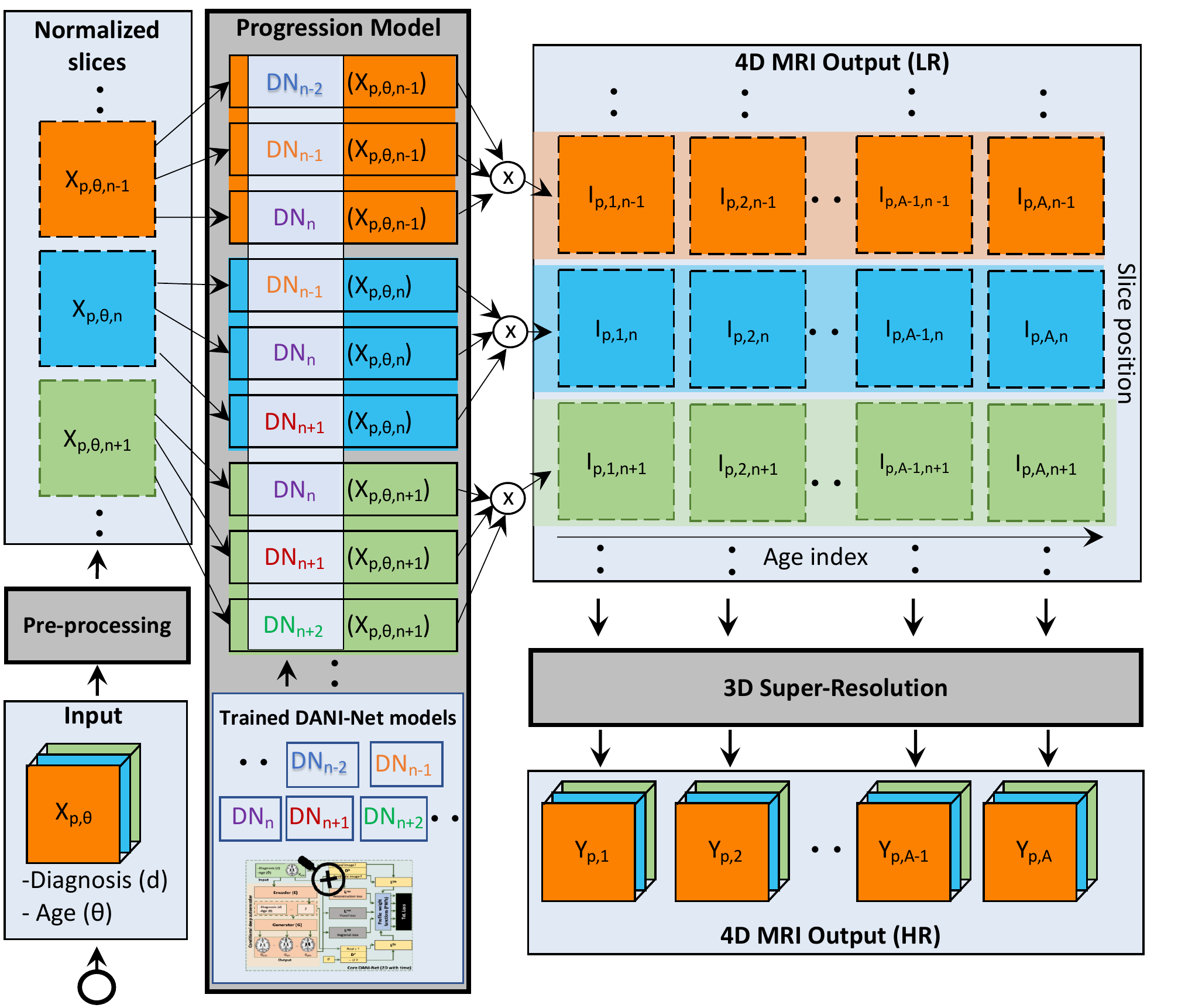}
\end{center}
   \caption{The full 4D-DANI-Net pipeline consisting of three main blocks, each depicted in grey: i) a pre-processing block, ii) a progression model consisting of a set of separate 2D DANI-Net modules trained with the proposed 3D consistency strategy called PWFs and iii) a 3D super-resolution block. The dashed blue boxes represent intermediate outputs of each block.\label{fig:DANI-Net}}
\end{figure*} 

Lack of individualisation precludes accurate modelling of individual trajectories because all the simulated MRI scans have the same, group-level deformation pattern. Approaches with this limitation usually create a spatiotemporal model that learns only one monotonic behaviour across all subjects~\cite{huizinga2018spatio,davis2010population,dalca2019learning,zhang2016consistent} or a few morphological templates associated to specific sub-groups~\cite{camara2006phenomenological,karaccali2006simulation,sharma2010evaluation,modat2014simulating}. An early attempt to overcome these restrictions exploited the power of deep generative models to propose a framework based on GANs which uses image arithmetic to combine atrophy patterns and manipulate MRI directly~\cite{bowles2018modelling}. However, this approach was restricted to linear (short-term) disease progression and was still based on learning group-level morphological changes that lose subject individuality over time. 

While solutions lacking individualisation do not completely fit the purpose of disease progression modelling,~\cite{vaden2020fully} have shown that sharing synthetic images reproducing group-level statistics is an alternative solution when it is not possible to share patient data due to privacy or data protection issues. 

The second and third limitations (poor image resolution and limited to 2D images) are mainly due to the computational cost required by a simulator. In fact, implementing effective methods for 3D, high-resolution brain images, often requires increased computational time due to memory issues~\cite{blumberg2018diqt}.  

One approach that suffers from these limitations is proposed in~\cite{khanal2017simulating} which combines a biophysical model and a deformation field obtained by non-rigid registration of two real images. This approach is constrained by memory restrictions that result in a trade-off between image resolution/dimensionality (e.g., 3D vs 2D), computation time and, ultimately precludes the utility of such an approach from scaling up to large, high-resolution datasets. Beyond the prohibitive computational cost,~\cite{khanal2017simulating} also relies on an atrophy lookup table rather than learning atrophy patterns from the data. 

Reducing the dimensionality from 3D to 2D MRI can ameliorate some of the computational limitations. For example, the simulator in~\cite{pathan2018predictive} proposed a predictive regression model for only 2D images. Instead of directly predicting images, this model predicts a vector momentum sequence~\cite{singh2013vector} associated with a baseline image where a Long Term-Short Memory (LSTM) network is used to encode the time-varying changes in the vector-momentum sequence, and a Convolutional Neural Network (CNN) is used to encode the baseline image of the vector momenta. ~\cite{xia2019consistent,xia2019learning} instead proposed a GAN-based adversarial training, which aimed to learn an age-based progression model for 2D slices of brain MRI scans. Our own preliminary work introduced a GAN-based framework, still only for 2D MRI~\cite{DANI-Net}, which was inspired by a face-ageing model~\cite{zhang2017age}.

Here, we introduce a framework to address all these limitations. We decompose the 4D problem (3D plus time) into learning multiple separate (2D plus time) models based on the slice-wise framework presented in~\cite{DANI-Net}. These separate models are unified using a new 3D training consistency strategy called Profile Weight Functions (PWFs) that preserves spatiotemporal continuity between 2D models. This memory-efficient strategy allows us to overcome limitation iii) -- restricted to 2D images, whereas a 3D super-resolution block is used to overcome limitation ii) -- poor image resolution. Lastly, we use a transfer learning strategy to obtain model individualisation to overcome limitation i) -- lack of individualization.


\begin{figure*}[t]
\begin{center}
    \includegraphics[width=0.65\textwidth,trim={0cm 5.8cm 17cm 0cm},clip]{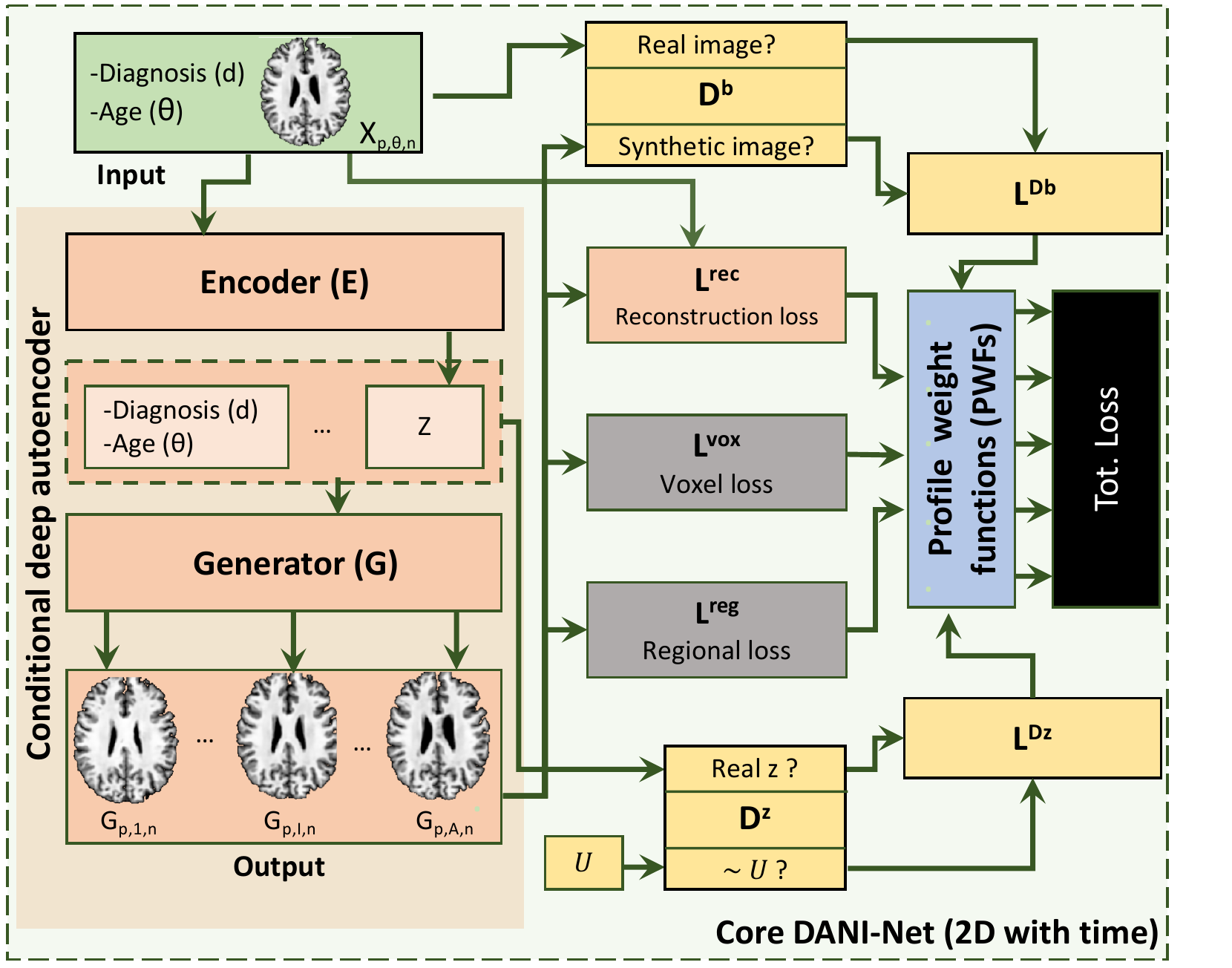}
\end{center}
   \caption{Single slice DANI-Net module used inside the propose framework 4D-DANI-Net. Each component of this module is identified by a different colour.\label{fig:DANI-Net-core}}
\end{figure*} 

\section{Methods}\label{sec:baseline}

4D-DANI-Net is a deep learning framework for synthesising high-resolution, longitudinal, subject-specific MRI scans. The core of the framework is a progression model based on adversarial training which includes biologically-informed spatiotemporal constraints to model neurodegeneration in ageing and dementia. Formally, 4D-DANI-Net generates the MRI sequence $Y_{p,i}$ with $i \in \{1...A\}$ representing the simulated series of $A$ time points for the subject $p$, initialised from a single input MRI $X_{p,\theta}$ acquired at age $\theta \in {\mathbb R^+}$. 

Our framework consists of three main blocks depicted in Fig.~\ref{fig:DANI-Net}: i) pre-processing; ii) progression model; and iii) 3D super-resolution. Pre-processing removes irrelevant variations in the data. Progression modelling is performed slice-wise (2D plus time) with 3D training consistency, as a set of DANI-Net models $DN_n$, where $n \in \{1...T\}$ represent different slice positions. Finally, our super-resolution block is a function that maps the resulting set of $T$ lower-resolution image slices $I_{p,i,n} \in \mathbb{R}^2$ for subject $p$ and time point $i$ obtained from each $DN_n$, to the high-resolution MRI $Y_{p,i} \in \mathbb{R}^3$. Below, we describe each block in detail.

\subsection{Pre-processing}
We use four pre-processing steps to prepare each input MRI $X_{p,\theta}$ for model training. This produces a set of $n$ normalized slices $X_{p,\theta,n}$ from each MRI. Samples with pre-processing failures were excluded from our experiments. 

The four steps are i) linear co-registration to 1mm isotropic MNI template using FLIRT-FSL~\cite{jenkinson2002improved}; ii) skull-stripping using BET-FSL~\cite{jenkinson2005bet2}; iii) extraction of the $n \in \{1...T\}$ axial slices from $X_{p,\theta}$; and iv) performing slice-wise intensity standardisation (zero mean, unit standard deviation). In combination, these steps reduce irrelevant variations in the data. Such variations can be caused by, e.g., scanner peculiarities and image orientation, which are irrelevant to the biological processes of interest.

\subsection{Progression Model}
For each axial slice in MNI space, we fit an independent 2D plus time progression model $DN_n$ (based on the original DANI-Net~\cite{DANI-Net}). Each DANI-Net model consists of three different sub-blocks (see Fig.~\ref{fig:DANI-Net-core}): a Conditional Deep Autoencoder (CDA) (coloured in pink); a set of adversarial networks (yellow); and a set of biological constraints (grey). We also introduce a novel PWFs strategy for unifying slice models into a 3D progression model during training (blue). 

\subsubsection{Conditional Deep Autoencoder (CDA)}
This block aims to learn a mapping between an initial manifold (representing brain MRI) and a lower-dimensional space, which we refer to as the latent space. This latent space is conditioned on other factors associated with the subject (i.e., current diagnosis, age) to allow manipulation of the image prediction on the original manifold according to these metadata.

More specifically, this block is composed of two deep neural networks: an encoder $E$ that embeds $X_{p,\theta,n}$ in a latent space $Z$, and a generator $G$ that projects samples in the latent space, back to the original manifold.
The latent vector $z$ is conditioned on two variables: $d \in \mathbb{N^+}$ --- a numerical representation [$0-3$] of diagnosis (cognitively normal CN, subjective memory concern SMC, early/late mild cognitive impairment E/LMCI, Alzheimer's disease AD); and $a \in \mathbb{N^+}$ --- an age index binned into $A$ groups. This age binning allows learning of morphological changes between age groups and prevents the CDA from memorizing (in the latent space) the age $\theta$ as an individual representation for each sample and thereby overfitting to age.

The CDA is trained using a reconstruction loss $L^{\textrm{rec}}_{p,n}$ that minimizes the difference between the input $X_{p,\theta,n}$ at age $\theta $ and the output sequence $G_{p,i,n} = G(E(X_{p,\theta,n}), i, d)$ with $i \in \{1...A\}$. This difference is weighted using a fuzzy Gaussian membership function $\mu_{i}[m_i,\sigma_i]$ centred on the average age $m_i$ of each age bin, with width $\sigma_i\propto\sqrt{\delta_i}$ proportional to the maximum age difference $\delta_i$ inside each bin. This preserves similarity between the input and the generated sequence, weighting nearer ages more heavily. Formally, $L^{\textrm{rec}}_{p,n}$ is described as follows:
\begin{equation}\label{L_rec_new}
L^{\textrm{rec}}_{p,n}=\sum_{i=1}^{A}L_{2}\big(X_{p,\theta,n} , G_{p,i,n} \mu _{{i}}[m_i,\delta_i] \big).
\end{equation}

\subsubsection{Adversarial Training}
GANs are a class of adversarial deep neural networks that have been successfully used to generate high-quality images across a wide range of tasks.  

We introduce a new adversarial training technique for the 4D-DANI-Net. In our case, the generator network $G$ (the decoder of our CDA) learns how to create synthetic realistic brain images. Simultaneously, we use two discriminators, $D^z$ and $D^b$, trained adversarially with the encoder $E$ and the decoder $G$ of our CDA.

More specifically, $G$ is trained to fool $D^b$, i.e., to generate brain MRI with a similar distribution to the initial true distribution. Simultaneously $D^b$ is trained to discriminate between empirical and synthetic brain MRI (generated by $G$). To train $D^b$ we use the following loss function:
{\begin{equation}\label{dbrain_loss}
\min\limits_{G}\max\limits_{D^b}\mathbb{E}_{p}\big[\log D^b\big(X_{p,\theta,n})\big) \big] + 
\mathbb{E}_{p}\big[1-\log D^b\big(G(E(X_{p,\theta,n}),a,d)\big)\big] ,
\end{equation}} 
where $\mathbb{E}$ is the expectation, $D^b$ estimates the probability that a slice contains a realistic brain and $E(X_{p,\theta,n})$ is the latent vector obtained from $X_{p,\theta,n}$. 

The second discriminator $D^z$ is trained adversarially with the encoder $E$, to produce $z$ with a uniform prior $\mathbb{U}$ and smooth temporal progression. To train $D^z$ we use the following loss function: 
{\begin{equation}\label{dz_loss}
\min\limits_{E}\max\limits_{D^z}\mathbb{E}_{z^*}\big[ \log D^z(z^*) \big]+\mathbb{E}_{p}\big[1- \log D^z(E(X_{p,\theta,n})) \big],
\end{equation}}
where $z^*$ is a vector sampled from $\mathbb{U}$ and $D^z$ estimates the probability that a vector comes from $\mathbb{U}$.

\subsubsection{Biological Constraints}\label{biological_constrain}
To capture the patterns of image intensity changes that accompany disease progression across time, 4D-DANI-Net uses two separate loss functions at different spatial scales: voxel-level $L^{\textrm{vox}}$ and region-level $L^{\textrm{reg}}$. These losses impose biological constraints that mimic neurodegeneration by ensuring monotonically decreasing intensity (brain tissue density~\cite{vemuri2010serial}) that is consistent with normal ageing and/or dementia.

For the synthetic output $G_{p,a,n}$ with $a$ equal to the bin index for age $\theta$, the voxel-level loss function $L^{\textrm{vox}}_{p,n}$ penalizes non-monotonic progression by imposing that all the voxels in $G_{p,i,n}$ with $i<a$ have equal or higher intensity, and that all the voxels in $G_{p,j,n}$ with $j>a$, have equal or lower intensity (recall that intensity is normalized in the first block of Fig.~\ref{fig:DANI-Net}).

$L^{\textrm{vox}}_{p,n}$ is defined as follows:
\begin{equation}\label{L_vox_new}
\begin{split}
L^{\textrm{vox}}_{p,n}=\frac{1}{2}\big[L_2(G_{p,a,n},\min(G_{p,1,n},...,G_{p,a-1,n}))+
\\
L_2(G_{p,a,n},\max(G_{p,a+1,n},...,G_{p,A,n}))\big]
\end{split}
\end{equation}

$L^{\textrm{vox}}_{p,n}$ models progression at the voxel level, but is incapable to model intensity changes that can occur at the global level (i.e., due to tissue deformation).

Therefore, we introduce a region-level loss function $L^{\textrm{reg}}_{p,n}$ that models slice-wise regional neurodegeneration through a set of pre-trained logistic regressors (LRs). Each regressor $LR_{n,q}$ is trained to predict intensity progression in fixed, overlapping region masks $q$. We describe how to generate these specific regions in Section~\ref{sec:Low-level-Consistency}. 

For slice $n$, the regressor takes three input features: age at baseline, age at follow-up, and diagnosis. We restrict each LR to train monotonically decreasing data by removing time-points where regional intensity increases (representing outliers). We also weigh the errors made by each $LR_{n,q}$ with the corresponding region size $s_{n,q}$, to induce consistent intensity within large regions. The contribution of $s_{n,q}$ helps to make this loss resistant to the noise in the MRI. 

Formally, $L^{\textrm{reg}_{p,n}}$ is defined as follows:
\begin{equation}\label{L_reg_new}
    \begin{split}
    L^{\textrm{reg}}_{p,n}&=\frac{1}{R_n(A-1)} \cdot  \\
    & \cdot \displaystyle\sum_{q=1}^{R_n}\bigg[\displaystyle\sum_{o=1}^{a-1}\bigg(\textrm{LR}_{n,q}(o,a,d)-\frac{\sum^*[G_{p,a,n} \odot r_{n,q}]+\epsilon}{\sum^*[G_{p,o,n} \odot r_{n,q}]+\epsilon}\bigg)\sqrt(s_{n,q})+
    \\
    &\displaystyle\sum_{o=a+1}^{A}\bigg(\textrm{LR}_{n,q}(a,o,d)-\frac{\sum^*[G_{p,o,n} \odot r_{n,q}]+\epsilon}{\sum^*[G_{p,a,n} \odot r_{n,q}]+\epsilon}\bigg)\sqrt(s_{n,q})\bigg],
    \end{split}
\end{equation}
where $R_n$ is the number of regions; $r_{n,q}$ are the region masks; $\mathrm{LR}_{n,q}(o,a,d)$ is the corresponding intensity change predicted from the logistic regressor for age $a$, conditioned on diagnosis $d$, starting from the baseline age $o$; $\epsilon=0.1$ avoids numerical errors; $\odot$ is the matrix Hadamard product (element-wise multiplication); and $\sum^*$ is the sums over brain voxels.

\subsubsection{Total Loss}\label{sec:DANI-NetTotalLoss}

Each single-slice DANI-Net model $DN_n$ is computed on the slice position $n$ and is trained to optimize all the losses ($L^{\textrm{reg}}_{p,n}, L^{\textrm{vox}}_{p,n}, L^{D^b}_{n}, L^{D^z}_{n}, L^{\textrm{rec}}_{p,n}$) at the same time. We illustrate this in the black block of Fig.~\ref{fig:DANI-Net-core}. 

The total loss is the weighted sum
\begin{equation}\label{Tot}
L^{\textrm{tot}}_{n} = w^{\textrm{reg}} \cdot \sum_p L^{\textrm{reg}}_{p,n} + w^{\textrm{vox}} \cdot \sum_p L^{\textrm{vox}}_{p,n} + w^b \cdot L^{D^b}_{n} + w^z \cdot L^{D^z}_{n}  + w^{\textrm{rec}} \cdot \sum_p L^{\textrm{rec}}_{p,n}
\end{equation}
where $L^{D^z}_{n}= \mathbb{E}_{z^*}\big[ \log D^z(z^*) \big]+ \mathbb{E}_{p}\big[1- \log D^z(E(X_{p,\theta,n})) \big]$ and $L^{D^b}_{n}=\mathbb{E}_{p}\big[\log D^b\big(X_{p,\theta,n})\big) \big]+\mathbb{E}_{p}\big[1-logD^b\big(G(E(X_{p,\theta,n}),a,d)\big)\big]$ are the cross entropies obtained respectively by the discriminators $D^z$ and $D^b$, for the slice position $n$ over all subjects $p$.

The weights allow for framework customization, such as:
\begin{itemize}[leftmargin=*]
\item[--] increasing $w^{\textrm{reg}}$ increases the contribution of disease progression (the LRs);
\item[--] increasing $w^{\textrm{vox}}$ regularizes voxel intensity changes for flat regions, but may increase rigidity of brain structures;
\item[--] increasing $w^b$ increases model generalization at the cost to decrease favours qualitatively realistic brain images; 
\item[--] increasing $w^z$ reduces temporal smoothing to allow rapid progression, which can introduce temporal discontinuity;
\item[--] increasing $w^{\textrm{rec}}$ increases similarity across age, which diminishes progression learned by the LRs.
\end{itemize}
Some loss functions optimize concurrent tasks, so finding the optimal configuration for these weights is nontrivial. Our strategy to accomplish this is via PWFs, that we describe in the next section.

\begin{figure*}[t]
\begin{center}
    \includegraphics[width=0.95\textwidth,trim={0cm 8.5cm 13cm 0cm},clip]{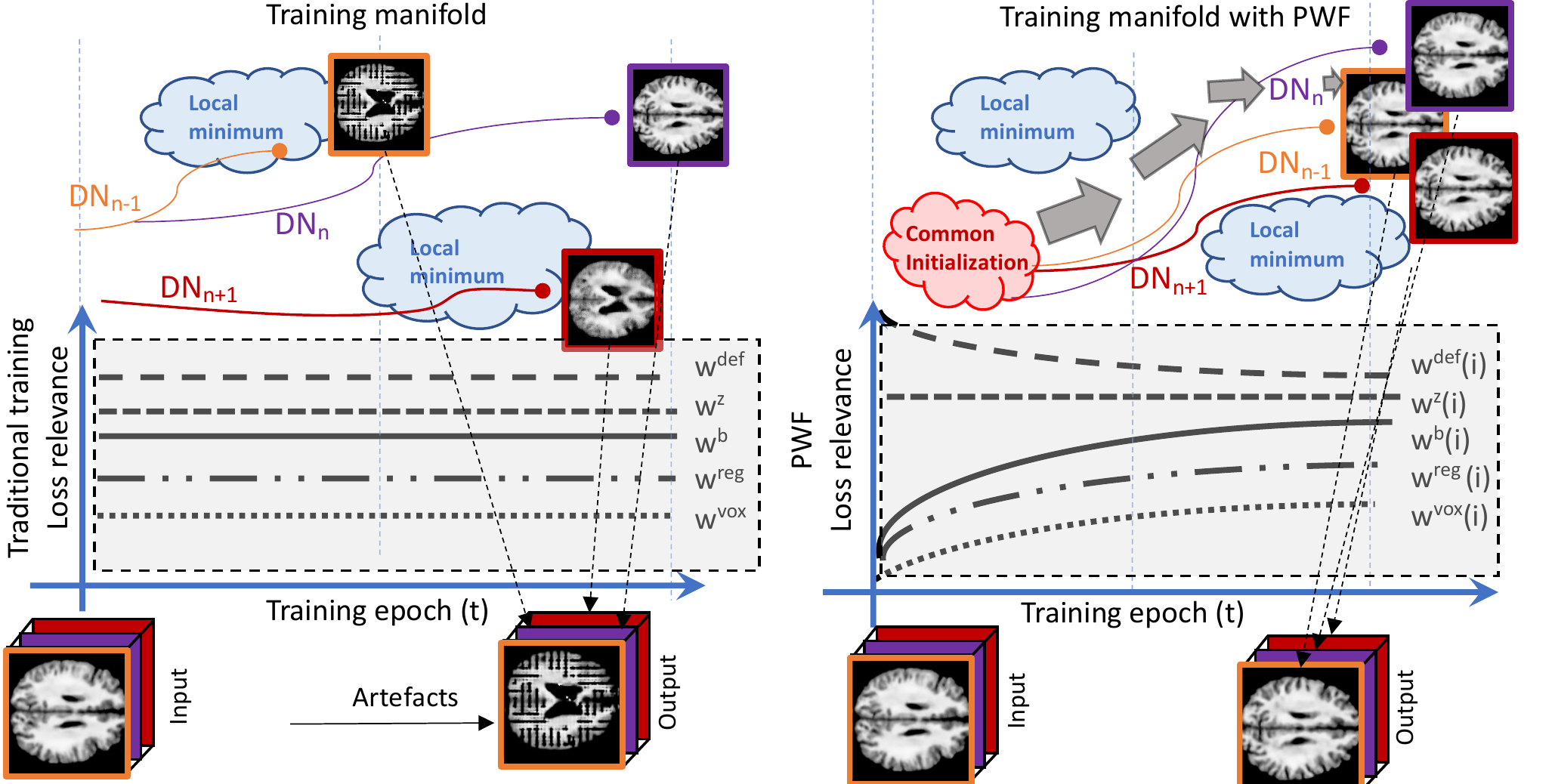}
\end{center}
\caption{Left: Training separated slice-based models using a complicated framework with multiple adversarial losses (i.e. 4D-DANI-Net) and each loss having a constant relevance along the entire training process, can lead to possible local minima. Right: Proposed PWFs strategy; the relevance of each loss during training is specified by profile functions with parameters learned via grid search. In this case, the training follows a specific path in the manifold and avoids local minima.}
\label{fig:weigth_profile}
\end{figure*} 

\subsubsection{Profiling Weight Functions (PWFs) for 3D Training Consistency}\label{sec:profile_weight_section}
In this section, we introduce PWFs that propose a way to dynamically weigh our five losses and unifying the training of the 2D slice-wise models~\cite{DANI-Net} in a computationally efficient manner.

Due to the complexity and non-convexity of our total loss function, training each $DN_n$ might be unstable. This is particularly problematic since convergence failures in a slice will generate spatial inconsistency artefacts in the synthetic 3D MRI. This is compounded by the adversarial components of DANI-Net ($D^z$ and $D^b$), as GANs are known to be prone to training instability~\cite{gulrajani2017improved,heusel2017gans}. 

The left block of Fig.~\ref{fig:weigth_profile} shows a hypothetical example that would create problems with classical adversarial training, as the competitor networks may reach a local minimum of the training manifold. 

To overcome this type of instability, the PWFs will guide training. It is inspired by a multistage learning strategy where humans solve a complex visual problem, i.e., optimizing simpler sub-tasks first. Explicitly, PWFs guide the system to focus on fewer loss functions at a time, i.e., providing greater regularization. This is achieved by dynamically weighting each component loss during every training epoch $t$. To do so, we use the following mean-reverting exponential function:
\begin{equation}
f(t)=\varrho^t \cdot b_{\textrm{loss}} + (1-\varrho^t) \cdot b_{\textrm{loss}} v^u 
\end{equation}
with parameters ($b_{\textrm{loss}}$, $v$ and $u$) optimized by a random search strategy on a grid, and measuring training convergence using the $L^{tot}_{n}$ on a validation set. The right side of Fig.~\ref{fig:weigth_profile} depicts how PWFs help to avoid local minima and, in our case, ensure that different models avoid the spatial mismatch that can cause image artefacts.

The final step for maintaining 3D consistency between consecutive slices is to smooth slice-wise models using a Gaussian-weighted ($\sigma=1.5$) average, that includes the $\pm2$ nearest-neighbour slices.

The workflow of the proposed progression model block is described schematically in Fig.~\ref{fig:DANI-Net}.

\subsection{3D Super-resolution}\label{3DSR:sec}
To recover lost anatomical detail due to the Gaussian smoothing (described in the preceding section), we include a 3D super-resolution block at the end of our pipeline (see Fig.~\ref{fig:DANI-Net}). This is based on a modified 3D densely-connected super-resolution network~\cite{chen2018brain} that uses pairs of low-resolution (LR) and high-resolution (HR) MRI for training a deep super-resolution neural network. 

We train this super-resolution block separately from the rest of our framework. To do so, we use as HR images the $X_{p,\theta,n}$ available in the training set, and as LR counterparts, the output obtained from the same input $X_{p,\theta,n}$, at the same age $\theta$ computed from our framework when the super-resolution block is disabled. 

Once these pairs of LR/HR are created, we use them to train the super-resolution network. We then proceed by attaching this trained SR-network onto the backbone of our trained system. This allows us to super-resolve the stacked output $I_{p,i,n}$ (generated by each $DN_n$) and to produce the required high-resolution MRI $Y_{p,i}$.

\subsection{Post-processing Using Transfer Learning}

At the inference stage, we use transfer learning to personalise and fine-tune the model on each test input brain. The aim of this procedure is to synthesize image evolution that reflects individual characteristics (here represented by brain morphology) and align each test MRI-slice to the model by age and diagnosis. 

\textcolor{black}{For each $DN_n$ and for each test subject $p$, we perform fine-tuning with an additional 50 training iterations on each single test input image, where only the MRI from each test subject's first visit is used. Here, only the parameters in the super-resolved block are frozen whilst all the other network parameters are fine-tuned to the specific morphology of the test individual's brain. The number of iterations required in the transfer learning step was determined empirically by visual inspection on the obtained images.}

\subsection{Additional Information: Region Extraction Based on Atlas}\label{sec:Low-level-Consistency}
The region masks used to train 4D-DANI-Net are pre-defined by the brain atlas proposed in~\cite{varentsova2014development} and imposed on each MRI using linear registration. The regions are extracted from the axial slices of this atlas. Additionally, to increase subject-specific variability, each region is augmented by applying morphological operators of erosion and dilatation. After this process, for each slice position, an average of 88 regions are generated, and a total of 9113 regions are extracted on the entire MRI (average size of a region is 517 voxels). As explained in Section~\ref{biological_constrain}, these regions represent each $r_{n,q}$ and are used to train the logistic regressors $LR_{n,q}$ required to embed the regional ratio of intensity changes in the system.

\section{Dataset and Training Details}\label{sec:training}
Data used in the preparation of this article were obtained from the ADNI database (adni.loni.usc.edu). The ADNI was launched in 2003 as a public-private partnership, led by Principal Investigator Michael W. Weiner, MD. The primary goal of ADNI has been to test whether serial Magnetic Resonance Imaging (MRI), Positron Emission Tomography (PET), other biological markers, and clinical and neuropsychological assessment can be combined to measure the progression of Mild Cognitive Impairment (MCI) and early AD.

In our experiments, we selected 12386 pre-processed T1-weighted MRI scans from $N=1216$ participants in the ADNI dataset. The scans were obtained using different scanners and at multiple sites. Participants were aged between 63 and 87 years old, and 28\% were CN, 4\% have been diagnosed with subjective memory concern, 54\% with mild cognitive impairment and 14\% with AD. Each participant has on average 4.7 MRI spanning 3 years. We divided our dataset in train-set (MRI: 9852; participants: 876), validation set (MRI: 1251; participants: 170), and test-set (MRI: 1283; participants: 170). In the test-set, we make sure that participants have at least one follow-up visit two years after baseline, to allow sufficient time for observable neurodegeneration to occur or to be excluded.

The first step of our training procedure uses the random grid search on the validation set to find the optimal parameters for our PWFs. The full PWFs and related parameters obtained from this procedure are reported below. 

\begin{align*}
w^{\textrm{reg}}(t) & =\varrho^t \cdot b_{\textrm{reg}} + (1-\varrho^t) \cdot b_{\textrm{reg}} \cdot v^{d_{reg}}\\
w^{\textrm{vox}}(t)  &=\varrho^t \cdot b_{\textrm{vox}}  + (1-\varrho^t) \cdot b_{\textrm{vox}} \cdot v^{d_{vox}}\\
w^{b}(t)    &=\varrho^t \cdot b_{b} + (1-\varrho^t) \cdot b_{b}   \cdot v^{d_{b}}\\
w^{z}(t)    &=\varrho^t \cdot b_{z}  + (1-\varrho^t) \cdot b_{z}   \cdot v^{d_{z}}\\
w^{\textrm{rec}}(t)  &=\varrho^t \cdot b_{\textrm{rec}}  + (1-\varrho^t) \cdot b_{\textrm{rec}} \cdot v^{d_{rec}}\\
\end{align*}
\textcolor{black}{
where $t$ is training epoch, $\varrho=0.99$ determines how fast the profile functions converge,
$v=10$ controls initial and final conditions of the weight functions, $b_{\textrm{reg}},b_{\textrm{vox}}, b_{b},b_{z}, b_{\textrm{rec}}$ are the proposed parameters describing the ``shape'' of the slope of each function and $d_{reg}$, $d_{vox}$, $d_{b}$, $d_{z}$, $d_{rec}$ describe the ``direction'' of the slope (ascending or descending). The range of values used in the grid search is chosen so that the product between each weight and the average value of the corresponding loss is normalized in the range [0-1]. The best values obtained during the grid search, together with the average values of each loss (which come from one training run of our system with the PWF block disabled) are reported in Table~\ref{table:grid_search}. }

\begin{table*}[!ht]
\caption{\textcolor{black}{Table summarizing the parameters used in the grid search and the corresponding best values obtained during the optimization.}}
\label{table:grid_search}

\begin{center}
\resizebox{1\textwidth}{!}{

\begin{minipage}{0.5\textwidth}
\begin{tabular}{c|c|c|c|c}
\hline
\multicolumn{4}{c|}{Shape Parameters }& \multirow{2}{*}{Average Loss Value}\\
\cline{1-4}
Parameter & Best Value & Search Range & Grid Step  &\\
\hline
$b_{\textrm{reg}}$&1.25&	[1-2]  &0.25&0.26 \\
$b_{\textrm{vox}}$&1.25& 	[1-2]  &0.25&0.24 \\
$b_{b}$&0.002 &	[0.001-0.004]  &0.001&72.27 \\
$b_{z}$&0.05 & 	[0.05	0.1] &0.01&12.34\\
$b_{\textrm{rec}}$ & 100& [75	175]	&25	&0.003\\
\end{tabular}
\end{minipage}
\hspace{3cm}

\begin{minipage}{0.5\textwidth}
\begin{tabular}{c|c|c|c}
\hline
\multicolumn{4}{c}{Direction Parameters}\\
\hline
Parameter & Best Value & Search Range & Grid Step \\
\hline
$d_{reg}$ & 1& [-1	1]	&2	\\
$d_{vox}$& 1& [-1	1]	&2	\\
$d_{b}$& 1& [-1	1]	&2	\\
$d_{z}$& 1& [-1	1]	&2	\\
$d_{rec}$& -1& [-1	1]	&2	\\
\end{tabular}
\end{minipage}
}
\end{center}
\end{table*}

\textcolor{black}{
The exponent values obtained for each PWF describe the optimization strategy used to train our system. Respectively, a negative/positive exponent tells the optimizer to focus on the corresponding loss earlier/later in the training process.}

\textcolor{black}{
Our interpretation of the obtained values suggests that the training should first focus on $L^{\textrm{rec}}_{p,n}$ (reconstruction loss) to learn a simple progression model based on conditional morphological deformations.} As $w^{\textrm{rec}}$ decays, the other weights increase towards their asymptotes, with our parameter search results meaning that $w^{\textrm{vox}}$ and $w^{\textrm{reg}}$ dominate over $w^z$ and $w^b$. In practical terms, this means that our PWFs favour optimization of losses associated with biological considerations (voxel-wise and region-wise neurodegeneration) over and above temporal smoothing ($w^z$) and image realism ($w^b$). The converse (favouring spatiotemporal similarity over progression modelling) can result in overfitting to individual morphology, and poor generalization performance of the model when applied to new individuals. This is because, generated images might be quite different from the training samples and, in this case, the discriminator $D^b$ would recognize them as unrealistic although they could potentially be real. This behaviour can drive the generator $G$ to avoid creating these brain structures although they are reasonable, which amounts to overfitting of individual morphology.

Although traditional initialization techniques can be used within our pipeline (i.e. Xavier), we found better performance when all the models start from a common initialization (red cloud in Fig.~\ref{fig:weigth_profile}). In practice, this common initialization consists of a model pre-trained with only 100 iterations on central axial slices of the training images.

Once the PWFs are defined, we proceed by training our $T$= 95 $DN_n$ models, each associated with one of the different slices $n$. The number of time points $A$ for the age is fixed to 10. Output MRI having an intermediate age value within these fixed points are obtained by a weighted linear interpolation of the two closest MRI. 

The architectures of each network $E$, $G$, $D^b$, $D^Z$ are based on the implementation proposed in~\cite{zhang2017age}. The size of the latent space Z is fixed to 200.
Each $DN_n$ is trained using the same PWFs and the same training configuration that is based on the stochastic gradient descent solver, ADAM ($\alpha$ = 0.0002, $\beta$1 = 0.5). We stop the training procedure after 300 epochs where each iteration uses a random mini-batch with 100 slices having the size of 128$\times$128 pixels. 

\begin{figure*}[t]
\begin{center}
    \includegraphics[width=0.8\textwidth,trim={0cm 0.8cm 9cm 0cm},clip]{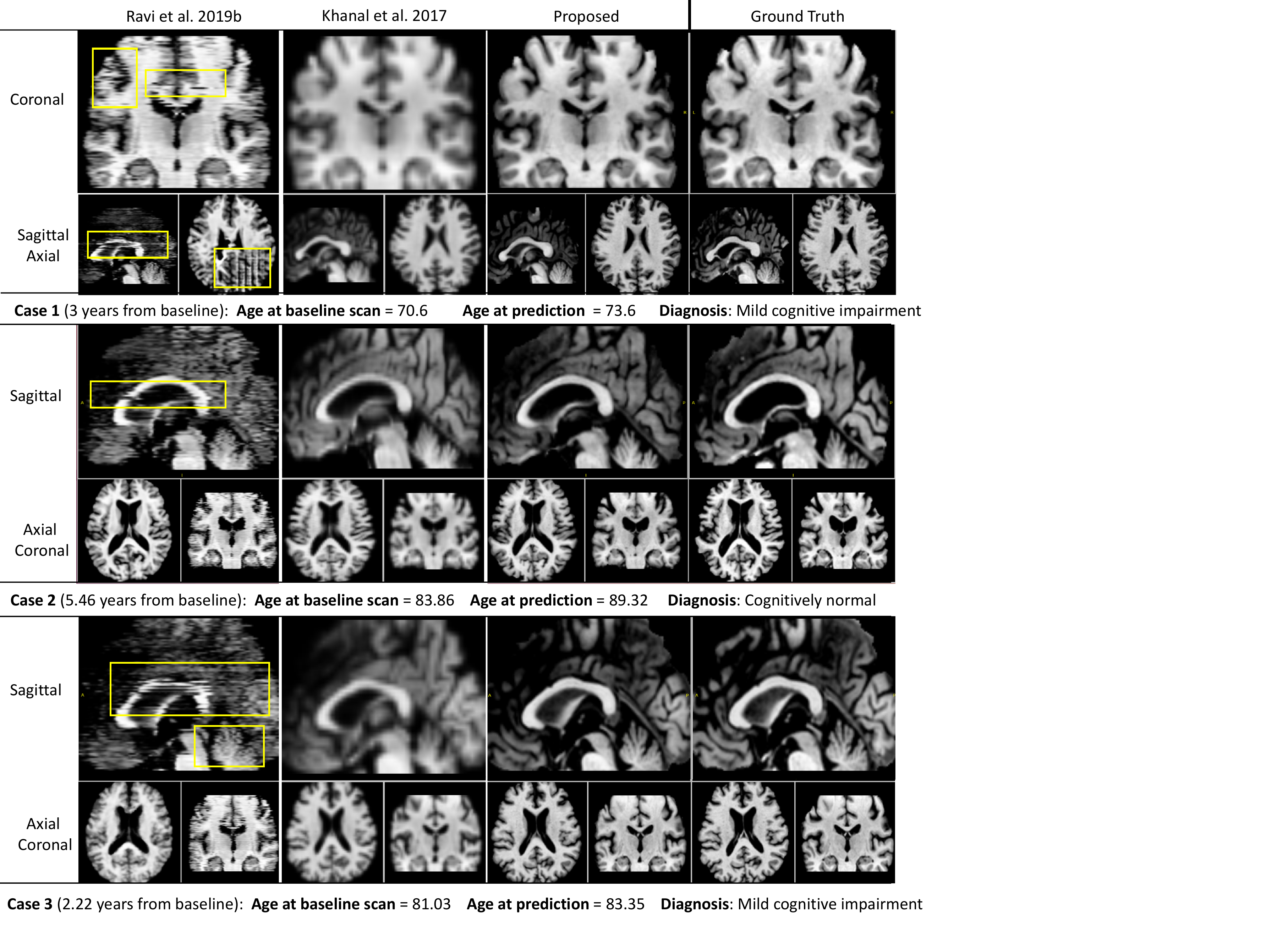}
\end{center}
   \caption{Qualitative comparison study: Synthetic MRI, generated starting from the baseline scan, for three representative test cases (rows) across different MRI simulator models (columns).
   }
\label{fig:visual_results_comparison}
\end{figure*}

\begin{table*}[!ht]
\caption{Quantitative comparison study: Mean absolute error ($\pm$  standard deviation) in predicted regional volumes of the brain, expressed as a percentage of total brain volume.}
\label{table:comparison}

\begin{center}
\resizebox{1\textwidth}{!}{
\begin{tabular}{c||c|c|c|c|c|c}

\multirow{2}{*}{Framework}&\multicolumn{2}{c|}{Small regions}&\multicolumn{4}{c}{Large regions}\\
\cline{2-7}

& \multicolumn{1}{c|}{Left Hippocampus } & Right Hippocampus & Peripheral Grey Matter & \multicolumn{1}{c|}{Ventricular CSF} & \multicolumn{1}{c|}{Tot. Grey Matter} & Tot. White Matter  \\

\hline
\cite{DANI-Net}& 
0.062 $\pm$ 0.052&
0.064 $\pm$ 0.049&
3.997 $\pm$ 1.805&
1.197 $\pm$ 0.755&
1.845 $\pm$ 1.379&
1.845 $\pm$ 1.379\\

\hline
SVR&            
0.029 $\pm$ 0.020&
0.032 $\pm$ 0.021&
1.432 $\pm$ 1.065&
0.688 $\pm$ 0.534&
1.553 $\pm$ 1.244&
1.557 $\pm$ 1.249\\
\hline

SVR$^*$&            
\textbf{0.028 $\pm$ 0.019}&
0.032 $\pm$ 0.020&
1.406 $\pm$ 1.041&
0.675 $\pm$ 0.538&
1.539 $\pm$ 1.198&
1.557 $\pm$ 1.201\\

\hline
LME&           
35.342 $\pm$ 18.198&
3.599 $\pm$ 2.595&
1.452 $\pm$ 0.999&
0.584 $\pm$ 0.420&
1.524 $\pm$ 1.053&
1.522 $\pm$ 1.068\\

\hline
LME$^*$&           
0.032 $\pm$ 0.024&
\textbf{0.030 $\pm$ 0.018}&
1.461 $\pm$ 1.009&
0.555 $\pm$ 0.415&
1.527 $\pm$ 1.059&
1.526 $\pm$ 1.061\\

\hline
Proposed &   
0.029 $\pm$ 0.028&
0.031 $\pm$ 0.031&
\textbf{0.771 $\pm$ 0.499}&
\textbf{0.257 $\pm$ 0.222}&
\textbf{0.829 $\pm$ 0.612}&
\textbf{0.829 $\pm$ 0.612}\\

\end{tabular}
}
\end{center}
\end{table*}

\section{Experiments and Results}\label{sec:result}
In our experiments, we first compare the proposed solution against state-of-the-art approaches using real follow-up as a ground truth. More specifically, we perform a qualitative assessment (Section~\ref{sec:qualitative_comparison}) based on the evaluation of image realism and artefacts, complemented with quantitative analyses (Section~\ref{sec:quantitative_comparison}) that measure the ability to generate MRI having accurate volumetric biomarkers. We then perform an ablation study (Section~\ref{sec:ablation_stady}) involving different configurations of 4D-DANI-Net to assess the contributions of each component block. Qualitative and quantitative assessments for the ablation study are presented respectively in Section~\ref{sec:ablation_stady_qualitative} and Section~\ref{sec:ablation_stady_quantitative}. We also evaluate the visual quality of our synthetic images via an evaluation survey (Section~\ref{sec:Clinical_Evaluation}) given to expert image readers, i.e., radiologists and neurologists. Finally, we present the computation time required for training and running our simulator in Section~\ref{sec:computational_time}, \textcolor{black}{and an experiment on model generalization using a new cohort in Section~\ref{sec:model_generalization}}.

\subsection{Qualitative Comparison Study}\label{sec:qualitative_comparison}
Here we compare our framework to the two state-of-the-art solutions available for MRI synthesis: i) the baseline DANI-Net obtained by independent training (and stacking) of 2D slice models~\cite{DANI-Net}; and ii) the biomechanical approach proposed in~\cite{khanal2017simulating}, which required down-sampling of the MRI resolution (by a factor of 2) for computationally feasible training times, followed by re-scaling to the original resolution using bilinear interpolation. 

Figure~\ref{fig:visual_results_comparison} shows that our approach provides the best results: fewer artefacts and superior resolution (less smoothing). Notably, images generated by~\cite{khanal2017simulating} show excessive smoothing, whereas images generated by~\cite{DANI-Net} contain notable artefacts.

\begin{figure*}[t]
\begin{center}
    \includegraphics[width=1\textwidth,trim={0cm 1cm 1cm 0cm},clip]{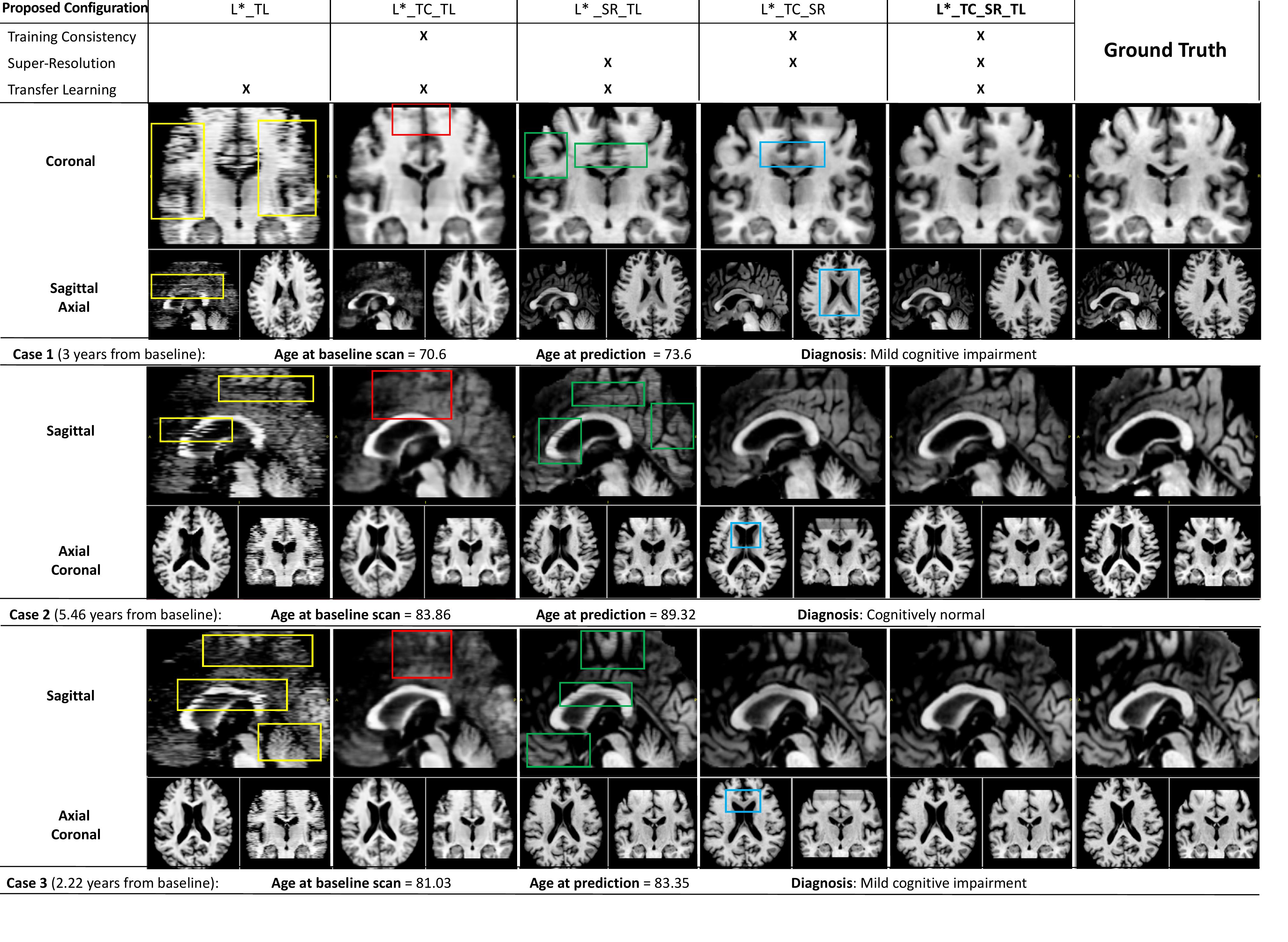}
\end{center}
   \caption{Qualitative ablation study: Synthetic MRI, generated starting from the baseline scan, for three representative test cases (rows) across different model configurations (columns) involving combinations of training consistency (TC), super-resolution (SR) and transfer learning (TL) blocks on top of the basic model L$^*$. Coloured boxes show: spatial discontinuity artefacts (yellow boxes) generated by unstable training; missing anatomical detail (red boxes) when super-resolution is not included; artefacts caused by super-resolution in the presence of spatial discontinuity artefacts (green boxes); and inaccurate morphology (blue boxes) in the ventricles when individualization is omitted from the model.}
\label{fig:visual_results}
\end{figure*}

\begin{table*}[t]
\caption{Quantitative ablation study: Mean absolute error ($\pm$ standard deviation) in predicted regional volumes of the brain, expressed as a percentage of total brain volume.}
\label{table:ablation}

\begin{center}
\resizebox{1\textwidth}{!}{
\begin{tabular}{c||c|c|c|c|c|c}

\multirow{1}{*}{Proposed framework }&\multicolumn{2}{c|}{Small regions}&\multicolumn{4}{c}{Large regions}\\
\cline{2-7}

configuration& \multicolumn{1}{c|}{Left Hippocampus } & Right Hippocampus & Peripheral Grey Matter & \multicolumn{1}{c|}{Ventricular CSF} & \multicolumn{1}{c|}{Tot. Grey Matter} & Tot. White Matter  \\

\hline
L$^*$\_TL &  
0.060 $\pm$ 0.049&
0.063 $\pm$ 0.058&
3.661 $\pm$ 1.751&
1.284 $\pm$ 0.794&
1.784 $\pm$ 1.213&
1.784 $\pm$ 1.213\\

\hline
L$^*$\_TC\_TL & 
0.060 $\pm$ 0.051&
0.060 $\pm$ 0.046&
2.396 $\pm$ 1.552&
1.236 $\pm$ 0.788&
1.761 $\pm$ 1.206&
1.761 $\pm$ 1.206\\

\hline
L$^*$\_SR\_TL &    
\textbf{0.029 $\pm$ 0.028}&
\textbf{0.031 $\pm$ 0.031}&
0.806 $\pm$ 0.539&
\textbf{0.250 $\pm$ 0.208}&
0.921 $\pm$ 0.685&
0.921 $\pm$ 0.685\\

\hline
L$^*$\_TC\_SR &
0.033 $\pm$ 0.027&
0.033 $\pm$ 0.028&
2.478 $\pm$ 1.270&
0.347 $\pm$ 0.275&
2.860 $\pm$ 1.427&
2.860 $\pm$ 1.427\\

\hline
L$^*$\_TC\_SR\_TL &  
\textbf{0.029 $\pm$ 0.028}&
\textbf{0.031 $\pm$ 0.031}&
\textbf{0.771 $\pm$ 0.499}&
0.257 $\pm$ 0.222&
\textbf{0.829 $\pm$ 0.612}&
\textbf{0.829 $\pm$ 0.612}\\

\end{tabular}
}
\end{center}
\end{table*}

\begin{table*}[!ht]
\caption{Quantitative ablation study: Percentage of improvements in framework accuracy for each component of our system.}
\label{table:ablation_C}

\begin{center}
\resizebox{1\textwidth}{!}{
\begin{tabular}{c||c|c|c|c|c|c|c}

\multirow{1}{*}{Considered framework }&\multicolumn{2}{c|}{Small regions}&\multicolumn{4}{c|}{Large regions}& Overall\\
\cline{2-7}

component& \multicolumn{1}{c|}{Left Hippocampus } & Right Hippocampus & Peripheral Grey Matter & \multicolumn{1}{c|}{Ventricular CSF} & \multicolumn{1}{c|}{Tot. Grey Matter} & Tot. White Matter  &  \\
\hline

Training Consistency (TC) &
         0.00\%&
         0.00\%&
   +4.34\%&
   -2.80\%&
   +9.98\%&
   +9.98\%&
   +3.58\%\\
\hline

Transfer Learning (TL) &
+12.16\%&
+6.06\%&
\textbf{ +68.88\%}&
  +25.93\%&
\textbf{  +71.01\%}&
\textbf{  +71.01\%}&
+42.50\%\\
\hline

Super-Resolution (SR) &
\textbf{ +51.66\%}&
 \textbf{+48.33\%}&
 +67.82\%&
 \textbf{ +79.20\%}&
 +52.92\%&
 +52.92\%&
\textbf{+58.80}\%\\

\end{tabular}
}
\end{center}
\end{table*}

\begin{table*}[!ht]
\caption{Quantitative ablation study: Percentage of improvements in framework accuracy for each term of our combined loss.}
\label{table:ablation_L}

\begin{center}
\resizebox{1\textwidth}{!}{
\begin{tabular}{c||c|c|c|c|c|c|c}

\multirow{1}{*}{Considered loss term}&\multicolumn{2}{c|}{Small regions}&\multicolumn{4}{c|}{Large regions}& Overall\\
\cline{2-7}

& \multicolumn{1}{c|}{Left Hippocampus } & Right Hippocampus & Peripheral Grey Matter & \multicolumn{1}{c|}{Ventricular CSF} & \multicolumn{1}{c|}{Tot. Grey Matter} & Tot. White Matter  &\\
\hline

\hline
Progression ($L^{\textrm{reg}}$ and $L^{\textrm{vox}}$) &
   +6.45\%&
    0.00\%&
  +14.71\%&
\textbf{+39.09\%}&
  +35.73\%&
  +35.73\%&
  +21.95\%\\
\hline

Reconstruction error ($L^{\textrm{rec}}$) & 
\textbf{+14.70\%}&
\textbf{+13.88\%}&
 +32.07\%&
  +19.68\%&
  +50.44\%&
  +50.44\%&
  +30.20\%\\
\hline

Realistic brain ($L^{D^b}$) &
-3.57\%&
   +6.06\%&
\textbf{+71.66\%}&
  +24.63\%&
\textbf{+72.62\%}&
\textbf{+72.62\%}&
+40.67\%\\

\hline
Temporal smoothing ($L^{D^z}$) &
  +12.12\%&
 +8.82\%&
+71.44\%&
  +25.29\%&
+71.71\%&
+71.71\%&
\textbf{+43.51\%} \\

\end{tabular}
}
\end{center}
\end{table*}

\subsection{Quantitative Comparison Study}\label{sec:quantitative_comparison}
Here we quantify the ability of the proposed 4D-DANI-Net to synthesize MRI that produce accurate regional volumes in the brain, as percentages of total brain volume (a standard approach to controlling for person-to-person variability in head size). Accuracy is presented as the mean and standard deviation in absolute error between synthetic and real images, across all 170 test cases. \textcolor{black}{This is achieved by applying brain segmentation in both simulated and synthetic MRI scans and computing volumes using the FSL library~\cite{FSL} for regions of interest relevant to ageing and Alzheimer's disease: left hippocampus, right hippocampus, peripheral grey matter, ventricular cerebrospinal fluid (CSF), total grey matter, and total white matter. }
\textcolor{black}{
Error for test subject $p$ in brain region $x$ is formulated as:}
\begin{equation}\label{Test_error}
     \begin{split}
     Err_{px}=\bigg|\frac{FSL(Y_{p},x)}{FSL(Y_{p},tb)}-\frac{FSL(Y^*_{p},x)}{FSL(Y^*_{p},tb)}\bigg|*100
     \end{split}
 \end{equation}
 
\textcolor{black}{where $Y^*_{p}$ is the real follow-up (ground truth), $Y_{p}$ is the simulated MRI, $FSL(Y_{p},x)$ is the estimated regional volume on $Y_{p}$ for the region $x$ obtained by the FSL library, and $FSL(Y_{p},tb)$ is the corresponding total brain volume.
}

Table~\ref{table:comparison} contains the results of quantitative comparison of our full model against other methods: DANI-Net~\cite{DANI-Net}, and a few other regression-based methods that have been used as benchmarks for predicting biomarker trajectories ~\cite{marinescu2020alzheimer}. Specifically, we consider a naive support vector regressor (SVR), a linear mixed-effects (LME) model, and two optimized regressor models, SVR$^*$ and LME$^*$, where 20\% of outliers were removed. Note that the regression-based models are trained directly on extracted brain volumes, with gender and diagnosis as covariates. These regressor approaches are incapable of generating simulated images.

For the LME model, we group the training set in four different groups based on diagnosis while the age and gender are considered both as random and fixed effects. For the SVR model, we used the RBF kernel with the hyper-parameters C=10 and coef0=0; and age, gender and diagnosis as predictive features. 

Apart from tweaking the baseline DANI-Net~\cite{DANI-Net} so that we could stack the different slices together and obtain the simulation on 3D MRI, we are unable to perform fair comparisons (same image resolutions) against other simulators (i.e.~\cite{khanal2017simulating}) due to the limitations presented in the introduction.

Table~\ref{table:comparison} shows that the worst-performing method is the original DANI-Net~\cite{DANI-Net}, which is not surprising because it was not designed for 3D MRI. 

The best performing method varies with brain region size. For large regions, 4D-DANI-Net (proposed approach) has the highest accuracy by a considerable margin: average reduction in error is $-33.2\%$ against SVR$^*$ and $-33.0\%$ against LME$^*$. For small regions, the SVR$^*$ and LME$^*$ slightly outperform 4D-DANI-Net. From this, we surmise that simple models are adequate for small regions, but are less capable to capture the complexity of neurodegeneration in larger regions.

In summary, from the comparison study, we can see that 4D-DANI-Net produces state-of-the-art performance for modelling neurodegeneration in ageing and Alzheimer's disease progression. 

\subsection{Ablation Study}\label{sec:ablation_stady}
In this section, we analyse the contribution of each component of our framework.

The configurations of 4D-DANI-Net considered in our ablation studies involve the basic model (denoted by L$^*$) obtained by independent training (then stacking together) of MRI slices, plus combinations of the 3D training consistency strategy (denoted by TC \textcolor{black}{and obtained when PWFs are used}), the super-resolution block (denoted by SR), and the transfer learning block (denoted by TL). See Section~\ref{sec:baseline} for details of each.

\subsubsection{Qualitative Ablation Study}\label{sec:ablation_stady_qualitative}
Our qualitative ablation study compared artefacts in synthetic images obtained by different configurations of 4D-DANI-Net for three representative test cases.

Figure~\ref{fig:visual_results} shows that the full configuration L$^*$\_TC\_SR\_TL produces visually superior synthetic MRI, i.e., fewer artefacts in comparison to synthetic MRI obtained by other configurations. In the approaches lacking 3D consistency constraints (L$^*$\_TL), the independent training of 2D slice-wise models leads to notable artefacts appearing in sagittal and coronal axes when networks do not converge (yellow boxes in Fig.~\ref{fig:visual_results}). As intended, such issues are almost eliminated through the use of our 3D training consistency strategy TC (L$^*$\_TC\_TL and L$^*$\_TC\_SR\_TL configurations). When TC is used without SR, anatomical details are often not visible (red boxes in Fig.~\ref{fig:visual_results}) and the images appear overly smooth. Conversely, when SR is used without TC, the super-resolution of artefacts introduces false structures (green boxes in Fig.~\ref{fig:visual_results}). Disabling the transfer learning procedure TL (configuration L$^*$\_TC\_SR) produces inaccurate morphology, i.e., excessive ventricles expansion, caused by lack of individualization (blue boxes in Fig.~\ref{fig:visual_results}).

For completeness, Fig.~\ref{fig:progression} shows an example of an entire simulation obtained using the full configuration of 4D-DANI-Net. Expected neurodegeneration is apparent in the sequence, including ventricular expansion, hippocampus contraction, and cortical thinning. 

\begin{figure*}[t]
\begin{center}
\includegraphics[width=0.95\textwidth,trim={0cm 2cm 8cm 0cm},clip]{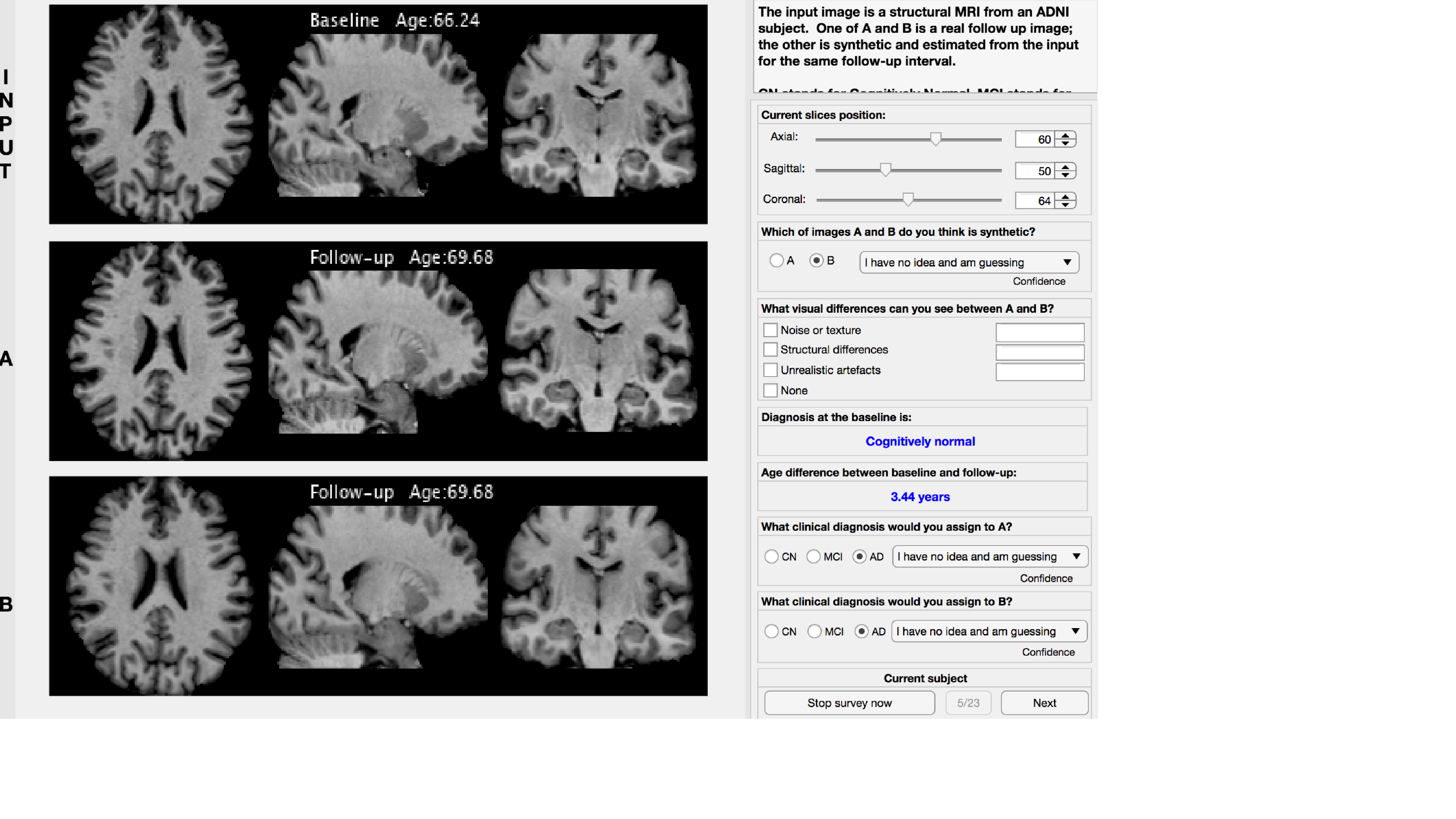}
\caption{Graphic User Interface used to perform the proposed survey.
\label{fig:survey-gui}}
\end{center}
\end{figure*}

\begin{figure*}[t]
\begin{center}
\includegraphics[width=1\textwidth,trim={0cm 5.4cm 0 0cm},clip]{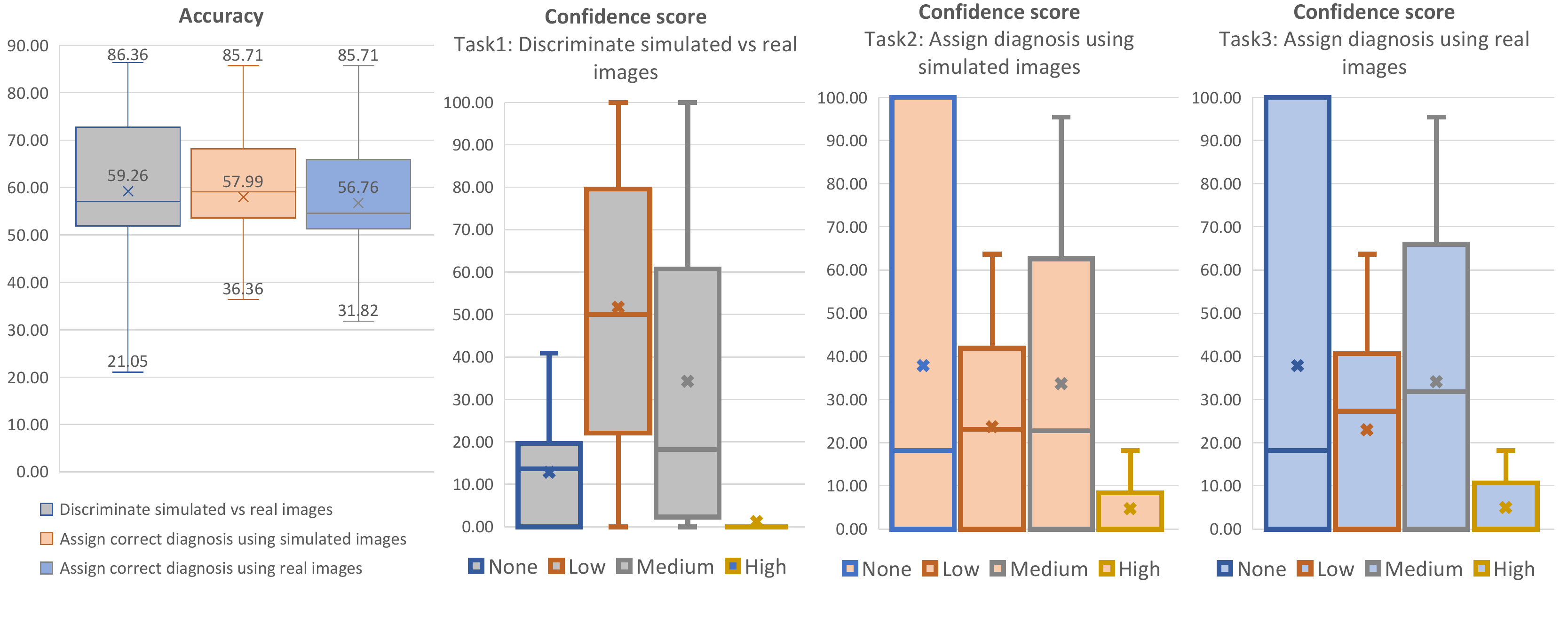}
\caption{Accuracy and confidence scores obtained from the participants of our survey on 3 different tasks: i) discriminating real images vs simulated images (bars in grey), ii) assigning diagnosis using simulated images (bars in orange) and iii) assigning diagnosis using real images (bars in blue).
\label{fig:survey1}}
\end{center}
\end{figure*} 

\begin{figure*}[t]
\begin{center}
\includegraphics[width=0.75\textwidth,trim={0cm 7cm 12cm 0cm},clip]{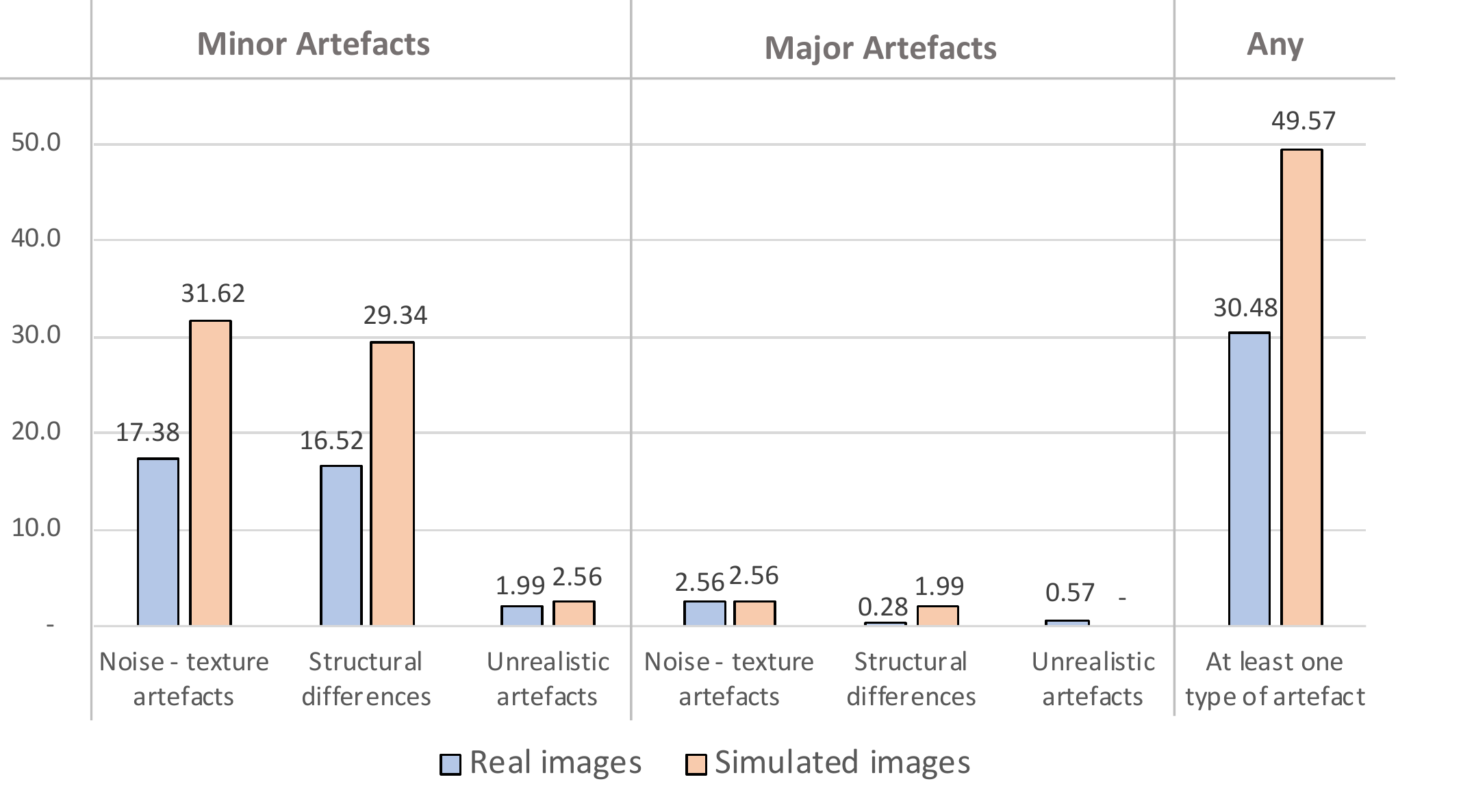}
\caption{Results on the participants' visual perception obtained during our survey. The considered artefacts are divided into 2 different severity scores (minor and major) and 3 different categories (noise/texture, structural differences and unrealistic artefacts). The results for the simulated images are in orange, whereas the results for the real images are in blue.
\label{fig:survey2}}
\end{center}
\end{figure*} 

\subsubsection{Quantitative Ablation Study}\label{sec:ablation_stady_quantitative}
Table~\ref{table:ablation} contains the results of our quantitative ablation study, which shows that the full model (L$^*$\_TC\_SR\_TL) produces the lowest absolute error in brain volume. Our 3D training consistency strategy TC reduces errors considerably: when TC is added to L$^*$\_TL, errors are reduced by an average (mean) of 7.6\%; when TC is added to the L$^*$\_SR\_TL configuration, errors are reduced by an average of 3.5\%.
Our super-resolution strategy SR improves accuracy significantly. In fact, SR was the largest contributor to accuracy by a considerable margin --- reducing errors by an average of 53.9\% when used with TL, and by 58.8\% when used with TL and TC. However, this last result also shows that super-resolution alone is not sufficient to maximize accuracy.

It is noteworthy that the absolute errors in gm and wm are identical, but they are in fact opposite in sign (not shown). This indicates that the source of volumetric errors is concentrated around the grey matter/white matter boundary, which is probably due to the well-known phenomenon of partial volume effects~\cite{weibull2008investigation}.

By looking at the results of the baseline~\cite{DANI-Net} in Table~\ref{table:comparison}, we note that any configuration of 4D-DANI-Net outperforms~\cite{DANI-Net}. Even the simplest configuration L$^*$\_TL reduces errors by an average of $2.8\%$. This is as expected since the baseline DANI-Net~\cite{DANI-Net} is similar to the simplest configuration of 4D-DANI-Net (L$^*$\_TL) except that the latter optimizes some of the loss functions, therefore providing better accuracy.

Table~\ref{table:ablation_C} summarizes the percentage of improvements in term of accuracy (error reduction) obtained when a specific component of our framework is included or excluded from the full configuration. Super-resolution provided the largest contribution (58.8\%) followed by the transfer learning (42.5\%) and the proposed training consistency (3.6\%).

Finally, Table~\ref{table:ablation_L} shows the percentage of improvements due to each term of our combined loss (L$^{tot}$). Temporal smoothing ($L^{D^z}$) provided the largest contribution (+43.5\%), closely followed by the adversarial loss related to brain realism ($L^{D^b}$, +40.7\%), then reconstruction error used to train the deep autoencoder ($L^{\textrm{rec}}$, +30.2\%) and disease progression modelling losses ($L^{\textrm{reg}}$ and $L^{\textrm{vox}}$, +21.9\%).

Our ablation studies cumulatively show that each component block and each loss of 4D-DANI-Net improves the performance in synthesizing a long-time sequence of personalised, high-resolution medical images with no discernible artefacts.

\subsection{Radiological Assessment of Visual Perception and Disease Stage}\label{sec:Clinical_Evaluation}
Finally, expert image readers evaluated simulated images against real images in terms of perceived visual artefacts as well as diagnostic accuracy.

To do so, we performed a survey where we recruited 21 participants (4 neurologists, 4 neuro-radiologists, 10 neuroimaging experts and 3 medical imaging researchers with an average of 9 years experience), and we asked them to evaluate 22 randomly selected cases extracted from the test set. From these cases, 3 out of 22 subjects have progressed in a different diagnosis during the follow-up scan whereas the remaining 19 subjects have maintained the initial diagnosis.

We set up an online web application (see Fig.~\ref{fig:survey-gui}) that shows, for each of the 22 cases, a T1-weighted brain MRI of a patient. Below this MRI scan, two more images labelled A and B shown in random order to avoid any selection bias. One of these 2 images is the real follow up MRI of the same initial subject; the other is the synthetic image generated starting from the initial MRI and obtained for the same follow-up interval. Each participant is asked to identify the simulated image in each of these 22 cases.

Additionally, during the survey, the participants were asked to identify and classify possible visual differences selected from 2 severity scores (minor and major) and 3 different categories (noise/texture, structural differences and unrealistic artefacts) and to assign a clinical diagnosis to both A and B in order to verify that there is no clinical inconsistency between real and synthetic images. The age and diagnosis at the baseline scan and the age at the follow-up scan were displayed to help the participants to assign the correct diagnosis. Finally, for each case and each different task in the survey, the participants were asked to provide a confidence score selected from the following list:
\small
\begin{itemize}
\item None: 'I have no idea and am guessing'
\item Low: 'I have low confidence'
\item Medium: 'I am reasonably confident' 
\item High: 'I am absolutely sure'
\end{itemize}
\normalsize

The results of this survey are presented in Fig.~\ref{fig:survey1} where each task is represented with a different colour: i) bars in grey depict the results related to the discrimination task (real vs simulated images), ii) bars in orange depict the results related to the diagnosis using simulated images and iii) bars in blue depict the results related to the diagnosis using real images.

On the first left graph of Fig.~\ref{fig:survey1}, we can see that participants can discriminate the real and simulated images with an accuracy of 59.3\%. We specifically noticed that neurologists achieved the highest accuracy (76.4\% $\pm$ 2.6\%), while neuro-radiologists obtained 68.0\% $\pm$ 7.1\%, neuroimaging experts 54.6\% $\pm$ 18.6\% and finally, medical imaging researchers 45.4\%$\pm$ 6.4\%.

\textcolor{black}{These results show that even the most highly trained participants have some difficulty discriminating between synthetic and real images (best score was 86.4\%), and all the participants together achieved just 59.3\% of accuracy that is slightly worse than the ideal case of random choice when the 2 classes are indistinguishable.}

In Fig.~\ref{fig:survey1} we can also see that the diagnosis using synthetic images is almost identical to the real follow-up (57.9\% vs 56.8\%) supporting the idea that our system is able to capture key aspects of disease progression.

In terms of the confidence scores related to the discrimination task (second left graph in Fig.~\ref{fig:survey1}), the majority of experts have select a low or a medium confidence score, confirming once again that the images cannot be easily discriminated.

For the confidence scores related to the assignment of the diagnosis (last two graphs in Fig.~\ref{fig:survey1}), these are distributed equally between the none and medium confidence scores and we did not find differences between the results on simulated image and the real ones.

In Fig.~\ref{fig:survey2} we report visual perception results from the survey of experts. These results show that the majority of the artefacts on the simulated images are minor noise/texture artefacts (31.6\%) and minor morphological structural differences (29.3\%). Only 2.6\% were minor unrealistic artefacts, 2.6\% major texture artefacts, 1.9\% major structural differences, and 0\% major unrealistic artefacts. From the results in this figure, we can also see that the simulated images have a slightly higher occurrence of artefacts with respect to the real images. In particular, in the last column, we can see that 30.5\% of real images have at least one artefact against 49.6\% for the synthetic images.
\par
$\\$
In conclusion, the highlights from our survey are as follows:
\small
\begin{itemize}
\item Simulated MRI scans contain minimal noise/texture artefacts and minor structural differences, approaching the levels of artefacts contained in real MRI scans.
\item Simulated MRI scans are diagnostically indistinguishable from real MRI scans.
\textcolor{black}{\item Simulated images and real MRI scans are not easy to discriminate (average performance is 59.3\%). However, experienced neurologists and neuro-radiologists were able to achieve reasonably high performance on this task (average 76.4\%).}
\end{itemize}
\normalsize

\subsection{Training and Inference Time}\label{sec:computational_time}
Having a cluster of GPUs with the total number of GPUs being similar to the number of slices in the MRI, allowed us to train each slice-based models in parallel. In our case, we have 50 NVIDIA GTX TITAN-X and 95 slices of MRI, and the total training time was approximately 3 days. The inference was much faster, in fact, on the same cluster, the computation time required to simulate the disease progression for a single MRI (including the transfer learning step) was in the order of a few minutes.

\begin{figure*}[t]
\begin{center}
\includegraphics[width=0.8\textwidth,trim={0cm 9.5cm 11cm 0cm},clip]{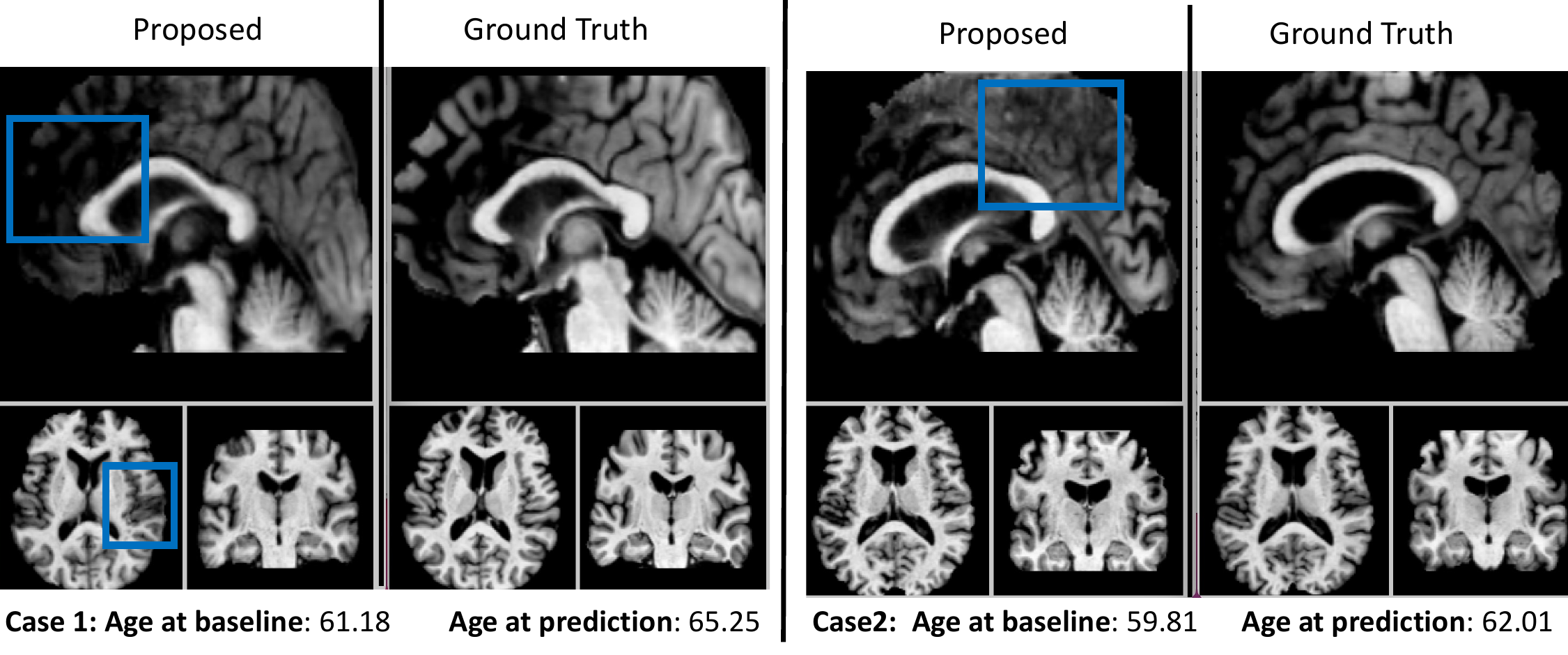}
\caption{\textcolor{black}{Two cases of model generalization issues causing blurred simulated images and occurring when the test subject is not well represented in the training set}. 
\label{fig:Extreme-Cases}}
\end{center}
\end{figure*}

\textcolor{black}{
\subsection{Model Generalization}\label{sec:model_generalization}
The design of our system allows simulating MRI scans from a new dataset without the need to retrain the system. However, for decent generalization, the current implementation of our framework requires the use of the same image modality (i.e., T1w-MRI) with a similar image resolution (i.e. 1mm isotropic). According to our framework design, the normalization step makes our model quite robust to change in scanner type or changes on the pre-processing pipeline whilst the personalization (transfer learning) step ensures good generalization to new subjects. To demonstrate this point, we have considered a second dataset called OASIS-3~\cite{lamontagne2019oasis} where we selected all the subjects diagnosed as CN or AD, aged in the range between 60-85 and having at least one follow-up 3 years after the baseline scan. After excluding all MRI scans where template registration and brain segmentation failed, we were able to evaluate 166 new subjects from this cohort. The results on this new dataset are presented in Table~\ref{table:generalization}.
}

\textcolor{black}{
Our results show that the model can generalize well to new data, with only slightly reduced performance on average. Generally speaking, AI model generalization is known to be a challenge, particularly when the training set is not fully representative of the target distribution. Indeed, reduced performance is expected here due to cohort differences, e.g., the distributions of age and follow-up duration, which differed between the ADNI and OASIS-3 datasets. We found that performance declined the most for large regions of the brain such as the cortical surface, which is well known to vary considerably between individuals, suggesting that personalization is the most challenging aspect of model generalization here. 
To demonstrate this, we show in Fig.~\ref{fig:Extreme-Cases} two cases where the model did not generalize well. A moderate drop in image quality is evident in large regions of the obtained images. In particular, in both cases of Fig.~\ref{fig:Extreme-Cases}, the images are blurred due to low numbers of individuals in the training set aged in the range between 60-62. Despite the performance reduction in such outlier cases, our model performed quite well in most cases. }

%

\textcolor{black}{
Lastly, our data-driven framework offers further potential to build generative models of other medical imaging modalities, e.g., tau PET in Alzheimer’s disease. This would require some methodological work such as modifying the biological constraints, used here to model neurodegeneration, to instead generate tau PET signal (for example).}

\begin{table*}[t]
\caption{\textcolor{black}{Results of our full pipeline on 2 different datasets.}}
\label{table:generalization}

\begin{center}
\resizebox{1\textwidth}{!}{
\begin{tabular}{c||c|c|c|c|c|c|c}

\multirow{1}{*}{Training}&\multirow{1}{*}{Testing }&\multicolumn{2}{c|}{Small regions}&\multicolumn{4}{c}{Large regions}\\
\cline{3-8}

cohort&cohort& \multicolumn{1}{c|}{Left Hippocampus } & Right Hippocampus & Peripheral Grey Matter & \multicolumn{1}{c|}{Ventricular CSF} & \multicolumn{1}{c|}{Tot. Grey Matter} & Tot. White Matter  \\

\hline
\hline
ADNI &ADNI &  
0.029 $\pm$ 0.028&
0.031 $\pm$ 0.031&
0.771 $\pm$ 0.499&
0.257 $\pm$ 0.222&
0.829 $\pm$ 0.612&
0.829 $\pm$ 0.612\\

\hline
ADNI &OASIS &
0.030 $\pm$ 0.029&
0.032 $\pm$ 0.031&
1.037 $\pm$ 0.735&
0.744 $\pm$ 0.509&
1.458 $\pm$ 0.926&
1.458 $\pm$ 0.926\\

\end{tabular}
}
\end{center}
\end{table*}

\section{Conclusion and Future Work}\label{sec:conclusion}
\textcolor{black}{
The aim of our system is to produce a ``digital twin’’ of the brain that can inform disease understanding and clinical decisions by predicting future evolution. Key clinical applications include: i) support earlier diagnosis by predicting future brain appearance of a specific subject; and ii) a personalised, virtual placebo for clinical trials. More specifically, for the later application, our experiments calculated the accuracy of future predictions in untreated individuals, which produces a practical baseline accuracy (with confidence intervals) for our system to be used as a virtual placebo. Any treatment result would have to exceed the confidence interval of the virtual placebo to prove to be effective. The concept of virtual placebos that we are proposing here can potentially revolutionise future clinical trials since this would avoid recruiting actual real placebo and cutting down the total cost of the clinical trial. In conclusion, our long-term vision for 4D-DANI-Net is to have a fully automatic system that can assess experimental treatments at the individual level, as well as suggesting the right dose to minimise side effects~\cite{yoon2018ganite}.}

In summary, in this paper, we presented a deep learning framework for brain image simulation in neurodegeneration, called 4D-DANI-Net, and demonstrated it in one of the biggest challenges of 21st-century healthcare: ageing and Alzheimer's disease. In particular, our work addresses a key gap in AI-enabled healthcare: generation of realistic and accurate synthetic medical images for model validation.  

Current state-of-the-art MRI simulators suffer three key limitations -- i) lack of individualization, ii) poor image resolution and iii) limited to 2D images -- that have precluded full 4D simulation of realistic and accurate high-resolution medical images, until now.

We addressed these limitations by introducing three memory-efficient components in our system. Firstly, the proposed profile weight functions control system instability \textcolor{black}{and although the parameters obtained in this work are ad hoc for this specific task, we believe that our PWF strategy can be a valid solution in many complex systems that suffer instability issues caused by optimizing simultaneously multiple adversarial networks. Therefore, such engineering novelty is important to ensure the stability of deep learning architectures with multiple networks, which are particularly complex in medical imaging, and therefore inherently unstable}. Secondly, the 3D super-resolution block is used to overcome low image resolution limitations. Thirdly, a new transfer learning strategy allowed us to personalise synthetic images for each individual.

We used quantitative and qualitative experiments to demonstrate the importance of each component of our pipeline and also compared our full framework against baseline models.

We see multiple exciting avenues for future work. Firstly, our framework can handle more advanced models of neurodegenerative disease progression and ageing, e.g., by conditioning on other factors such as demographics, lifestyle, and phenotype/genotype information for personalised medicine. This idea may be extended to investigate and test hypotheses of neurodegenerative disease mechanisms in a uniquely deep manner, which may help in the unsuccessful global efforts to develop effective treatments to date. Finally, and most importantly, our modular system can generalise beyond MRI and brain diseases to other medical imaging modalities, diseases, and organs of the body.

\section*{Acknowledgements}
The authors would like to thank NVIDIA Corporation for the donation of one of the GPUs used for this research. 

Data collection and sharing for this project was funded by the ADNI  (National Institutes of Health Grant U01 AG024904) and DOD ADNI (Department of Defense award number W81XWH-12-2-0012). ADNI is funded by the National Institute on Aging, the National Institute of Biomedical Imaging and Bioengineering, and through generous contributions from the following: AbbVie, Alzheimer's Association; Alzheimer's Drug Discovery Foundation; Araclon Biotech; BioClinica, Inc.; Biogen; Bristol-Myers Squibb Company; CereSpir, Inc.; Cogstate; Eisai Inc.; Elan Pharmaceuticals, Inc.; Eli Lilly and Company; EuroImmun; F. Hoffmann-La Roche Ltd and its affiliated company Genentech, Inc.; Fujirebio; GE Healthcare; IXICO Ltd.; Janssen Alzheimer Immunotherapy Research \& Development, LLC.; Johnson \& Johnson Pharmaceutical Research \& Development LLC.; Lumosity; Lundbeck; Merck \& Co., Inc.; Meso Scale Diagnostics, LLC.; NeuroRx Research; Neurotrack Technologies; Novartis Pharmaceuticals Corporation; Pfizer Inc.; Piramal Imaging; Servier; Takeda Pharmaceutical Company; and Transition Therapeutics. The Canadian Institutes of Health Research is providing funds to support ADNI clinical sites in Canada. Private sector contributions are facilitated by the Foundation for the National Institutes of Health (www.fnih.org). The grantee organization is the Northern California Institute for Research and Education, and the study is coordinated by the Alzheimer's Therapeutic Research Institute at the University of Southern California. ADNI data are disseminated by the Laboratory for Neuro Imaging at the University of Southern California. 

This project has received funding from the European Union's Horizon 2020 research and innovation programme under grant agreement No. 666992.

EPSRC grant EP/M020533/1 and the NIHR UCLH Biomedical Research Centre also supports this work.

NPO is a UKRI Future Leaders Fellow (MR/S03546X/1).

FB, NPO, and DCA are supported by the NIHR biomedical research centre at UCLH.

\bibliographystyle{unsrt}
\bibliography{refs}

\end{document}